\documentclass[aps,prd,reprint,groupedaddress,superscriptaddress,
floatfix,a4paper]{revtex4-1}

\usepackage{amsmath}
\usepackage{amssymb}
\usepackage{dcolumn}  
\usepackage{slashed}
\usepackage{graphicx}

\newcommand{\op}[1]{\mathcal{#1}} 
\def\firstmomentop{\op{O}_{\{\rho\mu\}}}
\def\MSbar{{\ensuremath{\overline{\mathrm{MS}}}}}
\def\latt{\mathrm{latt}}
\def\bigG{\frac{\alpha C_\mathrm{F}}{4\pi}}
\def\Mmf{M^{\mathrm{MF}}}
\def\wmf{w_0^{\mathrm{MF}}}
\def\Tmf{T_\mathrm{MF}}

\def\ODD{\op{O}_{DD}}
\def\Odd{\op{O}_{\partial\partial}}

\def\VDD{V_{DD}}
\def\Vdd{V_{\partial\partial}}
\def\cDD{c_{DD}}
\def\cdd{c_{\partial\partial}}

\newcommand{\bra}[1]{\langle #1|}
\newcommand{\ket}[1]{|#1\rangle}

\newcommand{\tpoL}{{\frac{2\pi}L}}
\newcommand{\ovra}[1]{\raisebox{0.09em}{$
\stackrel{\raisebox{-0.03em}{$\scriptstyle\to$}}{#1}$}{}}
\newcommand{\ovla}[1]{\raisebox{0.09em}{$
\stackrel{\raisebox{-0.03em}{$\scriptstyle\gets$}}{#1}$}{}}
\newcommand{\ovlra}[1]{\raisebox{0.09em}{$
\stackrel{\raisebox{-0.03em}{$\scriptstyle\leftrightarrow$}}{#1}$}{}}
\newcommand{\mres}{m_\mathrm{res}}
\newcommand{\PFM}{\langle\xi^1\rangle}
\newcommand{\PSM}{\langle\xi^2\rangle}
\newcommand{\VFM}{\langle\xi^1\rangle^\parallel}
\newcommand{\VSM}{\langle\xi^2\rangle^\parallel}
\newcommand{\VFMb}{\langle\xi^1\rangle^{\parallel\;\textrm{bare}}}
\newcommand{\VSMb}{\langle\xi^2\rangle^{\parallel\;\textrm{bare}}}
\newcommand{\tra}{\text{\tiny T}}

\newcommand{\rism}{\text{RI/SMOM}}
\newcommand{\ripm}{\text{RI$'$\kern-0.06667em/MOM}}
\newcommand{\bret}[1]{\left< #1 \right>}
\newcommand{\I}{\mathrm{i}}
\newcommand{\Tr}{\mathrm{Tr}}
\newcommand{\ms}{m_s}

\if@mathematic
\def\vec#1{\ensuremath{\mathchoice
    {\mbox{\boldmath$\displaystyle\mathbf{#1}$}}
    {\mbox{\boldmath$\textstyle\mathbf{#1}$}}
    {\mbox{\boldmath$\scriptstyle\mathbf{#1}$}}
    {\mbox{\boldmath$\scriptscriptstyle\mathbf{#1}$}}}}
\else
\def\vec#1{\ensuremath{\mathchoice
    {\mbox{\boldmath$\displaystyle#1$}}
    {\mbox{\boldmath$\textstyle#1$}}
    {\mbox{\boldmath$\scriptstyle#1$}}
    {\mbox{\boldmath$\scriptscriptstyle#1$}}}}
\fi

\newcommand{\eVdist}{\kern-0.06667em}

\newcommand{\Gev}{{\text{Ge}\eVdist\text{V\/}}}

\newcommand{\mev}{{\,\text{Me}\eVdist\text{V\/}}}
\newcommand{\gev}{{\,\text{Ge}\eVdist\text{V\/}}}

\begin{document}

\preprint{SHEP--10--20, DESY--10--110, CERN--PH--TH/2010--217}

\title{Lattice Results for Low Moments of Light Meson Distribution Amplitudes}



\author{R.~Arthur}
\email{r.arthur@sms.ed.ac.uk}
\author{P.A.~Boyle}
\email{paboyle@ph.ed.ac.uk}
\affiliation{SUPA, School of Physics, The University of Edinburgh, Edinburgh EH9 3JZ, UK}

\author{D.~Br\"ommel}
\email{d.broemmel@fz-juelich.de}
\altaffiliation[Current address ]{J\"ulich Supercomputing Centre,
                Institute for Advanced Simulation,
                Forschungszentrum J\"ulich GmbH, 52425 J\"ulich, Germany}
\affiliation{School of Physics and Astronomy, University of Southampton, Southampton SO17 1BJ, UK}

\author{M.A.~Donnellan}
\email{michael.donnellan@desy.de}
\affiliation{NIC/DESY Zeuthen, Platanenallee 6, 15738 Zeuthen, Germany}

\author{J.M.~Flynn}
\email{j.m.f{}lynn@soton.ac.uk}
\affiliation{School of Physics and Astronomy, University of Southampton, Southampton SO17 1BJ, UK}

\author{A.~J\"uttner}
\email{juettner@mail.cern.ch}
\affiliation{CERN, Physics Department, 1211 Geneva 23, Switzerland}

\author{T.D.~Rae}
\email{t.d.rae@phys.soton.ac.uk}
\author{C.T.C.~Sachrajda}
\email{cts@soton.ac.uk}
\affiliation{School of Physics and Astronomy, University of Southampton, Southampton SO17 1BJ, UK}

\collaboration{RBC and UKQCD Collaborations}
\noaffiliation

\date{\today}


\begin{abstract}
  As part of the UKQCD and RBC collaborations' $N_f = 2+1$ domain-wall
  fermion phenomenology programme, we calculate the first two moments
  of the light-cone distribution amplitudes of the pseudoscalar mesons
  $\pi$ and $K$ and the (longitudinally-polarised) vector mesons
  $\rho,\,K^*$ and $\phi$. We obtain the desired quantities with good
  precision and are able to discern the expected quark-mass dependence
  of SU(3)-flavour breaking effects. An important ingredient of the
  calculation is the nonperturbative renormalisation of lattice
  operators using the \ripm\ technique.
\end{abstract}

\pacs{}

\maketitle

\section{Introduction}
\label{sec:intro}

Light-cone distribution amplitudes (DAs) are important
nonperturbative quantities which (within the framework of collinear
factorisation) parameterise in partonic terms the components of the
hadronic wavefunction that control hard exclusive processes. Such
processes provide hadron structure information complementary to that
obtained from hard inclusive reactions.

The structure functions for inclusive processes are more accessible
both experimentally and theoretically owing to their larger cross
sections and branching ratios, simpler final-state detection and more
straightforward factorisation properties. They do not specify the
phases and correlations which would constitute amplitude-level hadron
structure information, but probe instead the bound states' partonic
content. Deep-inelastic scattering processes, for example, are
controlled by the charge and momentum of the struck parton and are
insensitive to its relation to the other hadronic constituents. The
associated parton distribution functions (PDFs) are therefore
single-particle probabilities, which reveal nothing about the role of
particular Fock states or of correlations between quarks and gluons.

Distribution amplitudes, involved in exclusive processes, always
appear in convolutions and, unlike the PDFs, are not directly
measurable. These exclusive processes are dominated by specific
partonic configurations. The outgoing quarks and gluons are unlikely
to form a given final-state hadron unless either they are
approximately collinear with small transverse separation, or one of
the partons carries almost all of the hadron's momentum (the soft
overlap or Feynman mechanism). In the former case, the basis for
collinear factorisation~\cite{Lepage:1980fj}, hard gluon exchange must
occur to allow the struck or decaying parton to communicate with the
others, turning them to the final direction. Since more partons
require more hard gluons, exclusive cross-sections and decay rates are
dominated by the valence Fock state at leading-order in $Q^2$, up to
soft effects.

Hard exclusive processes are therefore controlled at leading-order by
the distribution amplitudes of leading-twist (an operator's twist is
the difference between its dimension and its spin): essentially the
overlap of the hadronic state with the valence Fock state in which,
for a meson, the collinear quark-antiquark pair have small transverse
separation and carry longitudinal momentum fractions $u$ and $\bar{u}
= 1 - u$. The pion's electromagnetic form factor at large $Q^2$, for
example, can be written as a convolution of distribution amplitudes
$\phi_\pi(u,Q^2)$ for the incoming and outgoing pions with a
perturbatively-calculable hard-scattering kernel. Higher-twist DAs
associated with power-suppressed contributions originate in, for
example, higher Fock states~\cite{Ball:1998sk}. We consider only
leading, twist-$2$, DAs in this paper.

The phenomenological importance of hard exclusive processes has grown
since collinear factorisation was first established for cases such as
the pion's electromagnetic form factor and the $\gamma\gamma^*\pi$
transition form factor around 30 years
ago~\cite{Lepage:1980fj,Chernyak:1977fk,Efremov:1979qk,Farrar:1979aw}.
Of particular note is the theoretical description of hadronic $B$
decays, which have been studied in detail by BaBar and Belle and will
be studied by LHCb and at super-$B$ factories in order to constrain
the CKM matrix and to understand CP violation. Factorisation is more
difficult to establish in $B$-physics because the hard collinear and
soft mechanisms contribute at the same order in $1/m_b$. Two
approaches have been developed. In the QCD factorisation framework it
has been shown that collinear factorisation can be applied to leading
order in $1/m_b$ to a large class of nonleptonic
$B$-decays~\cite{Beneke:1999br, Beneke:2000ry, Beneke:2001ev}.
Soft-collinear effective theory (SCET)~\cite{Bauer:2000yr,
  Bauer:2001yt, Bauer:2002nz} aims to provide a unified theoretical
framework for the factorisation of both hard-collinear and soft
effects. In both cases, distribution amplitudes play an important role
as nonperturbative inputs in flavour physics.

In this paper we focus on the distribution amplitudes of the light
pseudoscalar and longitudinally-polarised vector mesons, since, as we
shall discuss in Sec.~\ref{subsec:moms}, their lowest moments are of
phenomenological interest and are calculable on the lattice. For
pseudoscalars, these quantities are relevant for decays such as $B
\rightarrow \pi\pi$ and $B \rightarrow \pi K$; they also appear in
light-cone sum rule (LCSR) expressions for the form factors of
semileptonic decays such as $B \rightarrow \pi l \nu$. For hard
exclusive processes involving the light vector mesons $\rho, K^*$ and
$\phi$, polarisation-dependence can reveal much about the underlying
dynamics, with the longitudinally- and transversely-polarised final
vector meson states often involving different aspects of weak
interaction physics~\cite{Braun:2003jg}. Examples are the exclusive
semileptonic $B \rightarrow \rho l \nu_l$, rare radiative $B
\rightarrow \rho\gamma$ or nonleptonic, e.g.\ $B \rightarrow \pi\rho$,
decays of $B$-mesons, which are important for extracting CKM matrix
elements.

\subsection{Definitions}
\label{subsec:defs}

Mesonic light-cone DAs are defined from meson-to-vacuum matrix
elements of quark-antiquark light-cone operators, which are non-local
generalisations of those used to define the decay constants. For
example, for pions
\begin{multline}
  \bra0\overline q_2(z)\gamma_\nu\gamma_5\op{P}(z,-z)q_1(-z)
  \ket{\pi(p)}_{z^2 =0} \equiv\\
  i f_\pi p_\nu \int_0^1 du\;e^{i(u-\bar u)p\cdot z}\phi_\pi(u,\mu)
\end{multline}
and for longitudinally-polarised rho-mesons
\begin{multline}
  \bra 0\overline q_2(z)\gamma_\nu \op{P}(z,-z)
  q_1(-z)\ket{\rho(p;\lambda)}_{z^2=0} \equiv\\
  f_\rho m_\rho p_\nu\frac{\varepsilon^{(\lambda)}\cdot z}{p \cdot z}
  \int_0^1 du\;e^{i(u-\bar{u})p.z}\phi^{\parallel}_\rho(u,\mu)\,,
\end{multline}
where
\begin{equation}
  \label{eq:pdef}
  \op{P}(z,-z)=\op{P}\,\exp\Bigl(-ig\int_{-z}^{z}dw^\mu
    A_\mu(w)\Bigr)
\end{equation}
is the path-ordered exponential needed to maintain gauge invariance,
$\mu$ is a renormalisation scale, $u$ is the momentum fraction of a
quark, $\bar u = 1-u$ and $\varepsilon^{(\lambda)}$ is the
polarisation vector for a vector meson with polarisation state
$\lambda$. The distribution amplitudes are normalised by
\begin{equation}
  \int^1_0 du\;\phi(u,\mu) = 1\,.
\end{equation}
The definitions above involve the pion and rho-meson decay constants
defined by
\begin{align}
\bra0\overline q_2\gamma_\mu \gamma_5 q_1\ket{\pi(p)} &=
 if_\pi p_\mu\,,\\
\bra0\overline q_2\gamma_\mu q_1\ket{\rho(p;\lambda)} &=
 f_\rho m_\rho \varepsilon^{(\lambda)}_\mu\,.
\end{align}
The vector meson decay constant, $f_\rho$, and its coupling to the
tensor current, $f^T_\rho$, are of interest in their own right and we
have previously calculated~\cite{Allton:2008pn} the ratios
$f^T_V/f_V$, for $V\in\{\rho,K^*,\phi\}$ as part of our domain-wall fermion
(DWF) phenomenology programme.

\subsection{Moments}
\label{subsec:moms}

Moments of light-cone DAs are defined by:
\begin{equation}
  \langle\xi^n\rangle_\pi(\mu) =
   \int_0^1 du\, \xi^n\,\phi(u,\mu)\,,
\end{equation}
where $\xi \equiv u - \bar u = 2u - 1$ is the difference between
the longitudinal momentum fractions.

Since the moments are obtained from matrix elements of local
operators~\cite{Brodsky:1980ny} we can study them using lattice QCD.
The light-cone matrix elements which define the DAs themselves are not
amenable to standard lattice techniques, since in Euclidean space the
light-cone has been rotated to the complex direction. By expanding the
non-local operators on the light cone, we obtain symmetric, traceless
twist-$2$ operators. With the following conventions for continuum
covariant derivatives,
\begin{equation}
  \ovra{D}_\mu = \ovra{\partial}_\mu+ig A_\mu , \quad
  \ovla{D}_\mu = \ovla{\partial}_\mu-ig A_\mu , \quad
  \ovlra{D}_\mu = \ovla{D}_\mu-\ovra{D}_\mu ,
\end{equation}
the expressions relating the moments of DAs to the corresponding local matrix
elements are:
\begin{widetext}
\begin{subequations}
\label{eq:dadef}
\begin{align}
  \bra{0}\overline{q}(0)\gamma_\rho\gamma_5\ovlra{D}_\mu s(0)\ket{K(p)}
  &=
  \PFM_Kf_Kp_\rho \,p_\mu \,, \label{eq:k1def}\\
  \bra{0}\overline{q}(0)\gamma_\rho\gamma_5\ovlra{D}_\mu\ovlra{D}_\nu
  q(0)\ket{\pi(p)} \,
  &=
  -i\PSM_\pi f_\pi p_\rho p_\mu p_\nu \,, \label{eq:pi2def}\\
  \bra{0}\overline{q}(0)\gamma_\rho\ovlra{D}_\mu s(0)\ket{K^*(p,\lambda)}
  &=
  \VFM_{K^*}f_{K^*}m_{K^*}\frac{1}{2}\left(p_\mu\varepsilon^{(\lambda)}_\nu
    + p_\nu\varepsilon^{(\lambda)}_\mu \right)\,, \label{eq:kstar1def}\\
  \bra{0}\overline{q}(0)\gamma_\rho\ovlra{D}_\mu\ovlra{D}_\nu q(0)
  \ket{\rho(p,\lambda)}
  &=
  -i \VSM_\rho \, f_\rho m_\rho
  \frac{1}{3}\left(\varepsilon^{(\lambda)}_\rho p_\mu p_\nu +
    \varepsilon^{(\lambda)}_\mu p_\nu p_\rho + \varepsilon^{(\lambda)}_\nu
    p_\rho p_\mu \right) \,. \label{eq:rho2def}
\end{align}
\end{subequations}
\end{widetext}
The operators in the matrix elements above are all to be considered
symmetric and traceless in the free Lorentz indices. Meson-meson
rather than meson-vacuum matrix elements of the same operators lead to
moments of generalised parton distributions (GPDs).

Recent analyses, especially those based on QCD sum rules, deal instead
with the Gegenbauer moments, which arise from a conformal
expansion~\cite{Braun:2003rp, Ball:2003sc}, in which the conformal
invariance of (classical) massless QCD is used to separate
longitudinal and transverse degrees of freedom, analogous to the
partial wave expansion in ordinary quantum mechanics. All dependence
on the longitudinal momentum fractions is described by orthogonal
polynomials that form an irreducible representation of the collinear
subgroup of the conformal group, SL(2,$\mathbb{R}$). The
transverse-momentum dependence is represented as the scale-dependence
of the relevant operators and is governed by renormalisation-group
equations. The different `partial waves', labelled by different
conformal spins, do mix but not to leading-logarithmic accuracy.
Conformal spin is thus a good quantum number in hard processes up to
small corrections of order $\alpha_s^2$.

The asymptotic $Q^2 \rightarrow \infty$ DA is known from perturbative
QCD: $\phi_\mathrm{as} = 6u\bar{u}$. For the leading-twist
quark-antiquark DAs that we are interested in, the conformal expansion
can then be conveniently written as:
\begin{equation}
  \phi(u,\mu) = 6u\bar{u}\biggl(1+
   \sum_{n=1}^\infty a_n(\mu)\,C_n^{3/2}(2u-1)\biggr) \,
\end{equation}
where $C_n^{3/2}$ are Gegenbauer polynomials. To one-loop order the
Gegenbauer moments, $a_n$, renormalise
multiplicatively~\cite{Ball:2003sc}:
\begin{equation}
  a_n(\mu) = a_n(\mu_0) \left( \frac{\alpha_s(\mu)}{\alpha_s(\mu_0)}
  \right)^{(\gamma_{(n)}-\gamma_{(0)})/\beta_0} \,.
\end{equation}
The one-loop anomalous dimensions are:
\begin{equation}
  \gamma_{(n)} = \gamma_{(n)}^\parallel =
  C_F\biggl(1 - \frac{2}{(n+1)(n+2)} +4\sum_{j=2}^{n+1}1/j\biggr) \,,
\end{equation}
where $C_F=4/3$. Since the moments are positive and increase with $n$,
the effects of higher-order Gegenbauer polynomials are damped at
higher scales as the DAs approach their asymptotic form. The conformal
expansion can thus be truncated. Quantities such as the pion's
electromagnetic form factor, for example, are given by convolutions in
which the kernels are slowly-varying and the strongly-oscillating
Gegenbauer polynomials are washed out. The same conclusion is reached
by considering, rather than the conformal expansion, the
diagonalisation of the ERBL
equations~\cite{Lepage:1979zb,Lepage:1980fj, Efremov:1979qk,
  Efremov:1978rn} which govern the evolution of the DAs much as the
DGLAP equations~\cite{Gribov:1972ri, Lipatov:1974qm,
  Dokshitzer:1977sg, Altarelli:1977zs} govern the evolution of PDFs.

We can obtain values for the Gegenbauer moments from lattice
simulations since the Gegenbauer moments are combinations of
ordinary moments of equal and lower order, e.g.:
\begin{equation}
  a_1 = \frac53\langle\xi^1\rangle, \quad
  a_2 = \frac7{12}\left(5\langle\xi^2\rangle - 1\right) \,.
\end{equation}

\subsection{Status}
\label{subsec:status}

In this section, we summarise what is currently known about
leading-twist light-meson distribution amplitudes. For mesons of
definite G-parity, there is a symmetry under the interchange $u
\leftrightarrow \bar{u}$ of the two momentum fractions. In these
cases, the distribution amplitude is an even function of $\xi = u -
\bar{u}$ and the odd moments therefore vanish. Thus, $\PFM_\pi$,
$\VFM_\rho$ and $\VFM_\phi$ all vanish, while $\PFM_K$ and
$\VFM_{K^*}$ are SU(3)-flavour breaking effects.

Since $\PFM_K$ is the average difference between the fractions of
longitudinal momentum carried by the strange and light quarks,
\begin{equation}
  \label{eq:define_a1K}
  \PFM_K(\mu)= \int_0^1 du (2u-1)\,\phi_K(u,\mu)=\langle 2u-1\rangle \,,
\end{equation}
we may expect from the constituent quark model that the sign of
$\PFM_K = \frac{3}{5}a_1^K$ is positive and this is indeed the case.
In fact, $\PFM_K$ is an important SU(3)-breaking parameter and is
relevant for predictions of $B$-decay transitions such as $B
\rightarrow K,\;K^*$~\cite{Braun:2006dg}. For example, a light-cone
sum rule analysis leads to~\cite{Khodjamirian:2003xk}:
\begin{equation}
  \frac{f^{BK}_+(0)}{f^{B\pi}_+(0)} =
  \frac{f_K}{f_\pi}(1 + c_1a^K_1) + \dots \,,
\end{equation}
where $f^{BP}_+(0)$ is the vector $B\to P$ form factor at zero
momentum transfer and $c_1 \sim O(1)$. Other examples include the
ratio of the weak radiative decay amplitudes $B \rightarrow
\rho\gamma$ and $B \rightarrow K^*\gamma$, where the main theoretical
error originates from such SU(3)-breaking effects. The measured ratio
of these decay rates allows for a determination of the ratio of CKM
matrix elements $|V_{ts}|/|V_{td}|$.

There have been three main approaches to the study of DAs: extraction
from experimental data, calculations using QCD sum rules and lattice
calculations. The overall normalisations are given by local hadronic
matrix elements, essentially the decay constants, which have already
been discussed and are partly accessible experimentally, and partly
have to be calculated theoretically. The shapes of the leading-twist
distribution amplitudes, in the form of the Gegenbauer moments, can be
determined from experiments by analysing data on form factors such as
$F_{\gamma\gamma^*\pi}$, which was studied by the CLEO
experiment~\cite{Gronberg:1997fj}, and the pion's electromagnetic
form factor, $F_\pi^\mathrm{em}$~\cite{Bakulev:2003gx}. There is a
lack of sufficiently accurate data, however, and it is difficult to
avoid contamination from other hadronic uncertainties and higher twist
effects. As a result, the existing experimental constraints are not
very stringent.

Moments of DAs, then, must largely be determined from theory.
Lattice~\cite{Braun:2006dg, Braun:2007zr, Braun:2008ur,
  Martinelli:1987si, Daniel:1990ah, DelDebbio:1999mq,
  DelDebbio:2002mq} and sum rule~\cite{Shifman:1978bx, Shifman:1978by,
  Chernyak:1983ej, Colangelo:2000dp} studies have usually focussed on
the second moment of the pion's distribution amplitude. However, the
early lattice results were largely exploratory while sum rule results
have an irreducible error of $\sim 20\%$ because it is not possible
properly to isolate the hadronic states.

The first moment of the kaon's distribution amplitude, for example,
has in the past been determined mainly from QCD sum rules, and
representative results include:
\begin{equation}
\label{eq:other}
a_1^K(1\gev)=
\begin{cases}
0.05(2) & \text{\cite{Khodjamirian:2004ga}}\\
0.10(12) & \text{\cite{Braun:2004vf}}\\
0.050(25) & \text{\cite{Ball:2005vx}}\\
0.06(3) & \text{\cite{Ball:2006fz}}
\end{cases}
\end{equation}
These results all have the expected sign, but the uncertainties are
around $50\%$. The reduction of such uncertainties is the chief
motivation of the lattice programme. In an earlier
publication~\cite{Boyle:2006pw,Boyle:2006xq}, we obtained
$\PFM_K(2\gev)\equiv 3/5 \,a_K^1\,(2\gev)=0.032(3)$. We note that in
addition to the UKQCD/RBC programme for the calculation of DA moments
on the lattice using $N_f=2+1$ domain-wall fermions, there is a
UKQCD/QCDSF programme using $N_f=2$ improved Wilson quarks
\cite{Braun:2006dg}. QCDSF have also published results for moments of
baryon DAs~\cite{Braun:2008ur}.
Lattice results for hadronic distribution amplitudes are considered in
a recent review of hadron structure from lattice QCD
in~\cite{Hagler:2009ni}.

The plan for the remainder of this paper is as follows. In
Sec.~\ref{sec:bare} we discuss the extraction of bare moments of
distribution amplitudes from Euclidean lattice correlation functions
(we use `bare' or `latt' to denote quantities before matching from the
lattice to the continuum). In Sec.~\ref{sec:numerics} we give the
details of our numerical calculations and present the bare results.
The renormalisation of those bare results is described in
Sec.~\ref{sec:renormalisation}. We then present our summary in
Sec.~\ref{sec:conclusion}.

\section{Bare Moments from Lattice Correlation Functions}
\label{sec:bare}

In this section, we describe our general strategy for the lattice
calculation of the unrenormalised lowest moments of light meson
distribution amplitudes. We obtain expressions for the first and
second moments $\PFM$ and $\PSM$ for pseudoscalar mesons and for the
longitudinal moments $\VFM$ and $\VSM$ for vector mesons, in terms of
Euclidean lattice correlation functions which can be computed by Monte
Carlo integration of the QCD path integral. In each case, we consider
a generic meson having valence quark content $\overline{q}_2q_1$,
where the subscripts indicate that the flavours of the two quarks may
be different. We will see below that we can obtain all of these
moments from ratios of two-point correlation functions and thus we
expect to benefit from a significant reduction of the statistical
fluctuations.

\subsection{Lattice Operators}

We now define the lattice operators used in the correlation functions
from which we extract the moments of the distribution amplitudes. We
use the following interpolating operators for the pseudoscalar and
vector mesons:
\begin{subequations}
\label{eq:interpolating}
\begin{align}
P(x) &\equiv \overline{q}_2(x)\gamma_5q_1(x), \\
V_\mu(x) &\equiv \overline{q}_2(x)\gamma_\mu q_1(x), \\
A_\mu(x) &\equiv \overline{q}_2(x)\gamma_\mu\gamma_5 q_1(x) \,.
\end{align}
\end{subequations}
Although we have written $P,\,V$ and $A$ as local operators in
Eq.~\eqref{eq:interpolating}, in the numerical simulations we use
smeared operators at the source of our correlation functions in order
to improve the overlap with the lightest meson states. Since the
effects of smearing cancel in the ratios constructed below, the
discussion in this section holds for both smeared and local
interpolating operators. We explain the details of our smearing
procedures in Sec.~\ref{subsec:simdet}. The operators in
Eqs.~\eqref{eq:dadef} from which the moments of the distribution
amplitudes are obtained are of course local operators.

In constructing the lattice operators of Eqs.~\eqref{eq:dadef}, we use
the following symmetric left- and right-acting covariant derivatives:
\begin{equation}
  \ovra{D}_\mu\psi(x) =
    \frac{1}{2a}\left[U(x,x{+}\hat\mu)
    \psi(x{+}\hat\mu)- U(x,x{-}\hat\mu)\psi(x{-}\hat\mu)\right] ,
    \label{eq:derdef}
\end{equation}
\begin{equation}
  \overline{\psi}(x)\ovla{D}_\mu =
    \frac{1}{2a}\left[\overline{\psi}
    (x{+}\hat\mu) U(x{+}\hat\mu,x) - \overline{\psi}(x{-}\hat\mu)
    U(x{-}\hat\mu,x)\right] ,
\end{equation}
where $U(x,y)$ is the gauge link going from site $x$ to site $y$ and
$\hat{\mu}$ is a vector of length $a$ in the direction $\mu$ ($a$
denotes the lattice spacing). The operators of interest are then
defined by
\begin{subequations}
\label{eq:ops}
\begin{align}
  \op{O}_{\{\rho\mu\}}(x) &\equiv
  \overline{q}_2(x)\gamma_{\{\rho}\ovlra{D}_{\mu\}}q_1(x) \,,\\
  \op{O}_{\{\rho\mu\nu\}}(x) &\equiv
  \overline{q}_2(x)\gamma_{\{\rho}\ovlra{D}_{\mu}\ovlra{D}_{\nu\}}q_1(x) \,,
  \\
  \op{O}^5_{\{\rho\mu\}}(x) &\equiv
  \overline{q}_2(x)\gamma_{\{\rho}\gamma_5\ovlra{D}_{\mu\}}q_1(x) \,,\\
  \op{O}^5_{\{\rho\mu\nu\}}(x) &\equiv
  \overline{q}_2(x)\gamma_{\{\rho}\gamma_5\ovlra{D}_{\mu}\ovlra{D}_{\nu\}}q_1(x) \,,
\end{align}
\end{subequations}
where the braces in the subscripts indicate symmetrisation of the
enclosed Lorentz indices, $\{\mu_1\dots\mu_n\} \equiv
\sum_{\mathrm{perms}\;s} \{\mu_{s(1)}\dots\mu_{s(n)}\}/n!$.

\subsection{Operator Mixing}
\label{subsec:opmix}

In the continuum the operators in Eq.~\eqref{eq:ops} transform as
second- or third-rank tensors under the Lorentz group. On the lattice
however we must consider their transformation properties under the
hypercubic group ${\cal H}_4$ of reflections and $\pi/2$ rotations,
together with the discrete symmetries parity $P$ and
charge-conjugation $C$, where the possibilities for operator mixing
are increased. A detailed study of the transformations of these
operators under ${\cal H}_4$ has been performed
in~\cite{Gockeler:1996mu}.

The choice of Lorentz indices in the operators used in simulations is
important both to keep the operator mixing simple and also to enable
the extraction of matrix elements using as few non-zero components of
momentum as possible. The latter is to avoid the associated
discretisation effects and statistical degradation.
$\op{O}_{\{\rho\mu\}}$ and $\op{O}_{\{\rho\mu\}}^5$ renormalise
multiplicatively under ${\cal H}_4$ when $\rho\neq\mu$. In the
notation of~\cite{Mandula:1983ut}, these operators transform under the
$6$-dimensional $6^{(+)}$ (for $\op{O}_{\{\rho\mu\}}^5$) or $6^{(-)}$
(for $\op{O}_{\{\rho\mu\}}$) irreducible representations of ${\cal
  H}_4$. The choice $\mu\neq\rho$ is the most convenient one for the
extraction of the first moment of the distribution amplitudes. Charge
conjugation symmetry combined with ${\cal H}_4$ ensures that there is
no mixing with operators containing total derivatives.

It is also possible to obtain the first moment from the four operators
$\op{O}_{\{\mu\mu\}}$ (or $\op{O}^5_{\{\mu\mu\}}$), which each
transform as four-dimensional reducible representations containing a
singlet. The three traceless operators transform as the
$3$-dimensional irreducible representation $(3,1)^{(+)}$ (without
$\gamma^5$) or $(3,1)^{(-)}$ (with $\gamma^5$). Subtracting the trace
involves the subtraction of a power divergence, so for the first
moment of the distribution amplitude of the $K$ and $K^*$ we avoid
this by evaluating the matrix elements of $\op{O}^5_{\{\rho\mu\}}$ and
$\op{O}_{\{\rho\mu\}}$ respectively with $\rho\neq\mu$.

Similarly for the second moment of the distribution amplitudes the
most convenient choice is to use $\op{O}^5_{\{\rho\mu\nu\}}$ or
$\op{O}_{\{\rho\mu\nu\}}$ with all three indices different, which
transform as the $(\overline{1/2,1/2})^{(+)}$ and
$(\overline{1/2,1/2})^{(-)}$ $4$-dimensional irreducible
representations respectively. Charge conjugation symmetry allows
mixing of $\op{O}^5_{\{\rho\mu\nu\}}$ and $\op{O}_{\{\rho\mu\nu\}}$
with operators containing total derivatives:
\begin{eqnarray*}
  \op{O}^5_{\{\rho\mu\nu\}}(x)
  &\text{ mixes with }
  &\partial_{\{\rho}\partial_\mu\,\left(\overline{q}_2(x)\gamma_{\nu\}}\gamma_5
    q_1(x) \right) \,,\\
  \op{O}_{\{\rho\mu\nu\}}(x)
  &\text{ mixes with }
  &\partial_{\{\rho}\partial_\mu\,\left(\overline{q}_2(x)\gamma_{\nu\}}
    q_1(x) \right) \,.
\end{eqnarray*}
The moments of the distribution functions are obtained from
non-forward matrix elements between a meson at non-zero four momentum
and the vacuum, so the total-derivative operators must be included in
the analysis.

\subsection{$\boldsymbol{\PFM_P}$ and $\boldsymbol{\PSM_P}$ from Correlation
  Function Ratios}

To obtain the first and second moments of the pseudoscalar meson
distribution amplitude, $\PFM$ and $\PSM$, we consider the following
two-point correlation functions:
\begin{subequations}
\label{eq:cpdef}
\begin{align}
  C_{A_\nu P}(t,\vec{p}) &\equiv \sum_{\vec{x}} e^{i \vec{p}\cdot\vec{x}}
  \bra{0} A_\nu(t,\vec{x}) P^\dagger(0) \ket{0} \,,\label{eq:capdef} \\
  C^5_{\{\rho\mu\}}(t,\vec{p}) &\equiv \sum_{\vec{x}} e^{i
    \vec{p}\cdot\vec{x}} \bra{0} \op{O}^5_{\{\rho\mu\}}(t,\vec{x}) P^\dagger(0)
  \ket{0} \,,\label{eq:c1pdef} \\
  C^5_{\{\rho\mu\nu\}}(t,\vec{p}) &\equiv \sum_{\vec{x}} e^{i
    \vec{p}\cdot\vec{x}} \bra{0} \op{O}^5_{\{\rho\mu\nu\}}(t,\vec{x}) P^\dagger(0)
  \ket{0} \,.\label{eq:c2pdef}
\end{align}
\end{subequations}
For a generic pseudoscalar meson $P$, we define $Z_P \equiv
\bra{P(p)}P^\dagger\ket{0}$ and the bare decay constant by
$\bra{0}A_\nu\ket{P(p)} \equiv ip_\nu f^{\textrm{bare}}_P$. The
operators $P^\dagger(0)$ in Eqs.~\eqref{eq:cpdef} are smeared as
explained below. At large Euclidean times $t$ and $T - t$, the
correlation functions defined above tend towards:
\begin{widetext}
\begin{align}
  C_{A_\nu P}(t,\vec{p}) &\rightarrow
  \frac{Z_P f^{\textrm{bare}}_P e^{-E_PT/2} \sinh((t{-}T/2)E_P)}{E_P} \,
  ip_\nu \,,\label{eq:C5r}\\
  C^5_{\{\rho\mu\}}(t,\vec{p}) &\rightarrow
  \frac{Z_P f^{\textrm{bare}}_P e^{-E_PT/2} \sinh((t{-}T/2)E_P)}{E_P} \,
  ip_\rho ip_\mu\PFM^{\textrm{bare}} \,,\label{eq:C5rm}\\
  C^5_{\{\rho\mu\nu\}}(t,\vec{p}) &\rightarrow
  \frac{Z_P f^{\textrm{bare}}_P e^{-E_PT/2} \sinh((t{-}T/2)E_P)}{E_P} \,
  ip_\rho ip_\mu ip_\nu \PSM^{\textrm{bare}} \,.\label{eq:C5rmn}
\end{align}
\end{widetext}
We can extract bare values for the first and second moments of the
pseudoscalar meson distribution amplitudes from the following ratios
of correlation functions:
\begin{subequations}
\label{eq:Pratio}
\begin{align}
  R^P_{\{\rho\mu\};\nu}(t,\vec{p}) &\equiv
  \frac{C^5_{\{\rho\mu\}}(t,\vec{p})}{C_{A_\nu P}(t,\vec{p})} \rightarrow
  i\frac{p_\rho p_\mu}{p_\nu}\PFM^{\textrm{bare}} \,,\label{eq:PFMratio} \\
  R^P_{\{\rho\mu\nu\};\sigma}(t,\vec{p}) &\equiv
  \frac{C^5_{\{\rho\mu\nu\}}(t,\vec{p})}{C_{A_\sigma P}(t,\vec{p})}
  \rightarrow -\frac{p_\rho p_\mu
    p_\nu}{p_\sigma}\PSM^{\textrm{bare}} \,.\label{eq:PSMratio}
\end{align}
\end{subequations}
Keeping in mind the operator mixing outlined above, we obtain the
first moment from $R^P_{\{\rho4\};4}(t,\vec{p})$ (the index $4$
corresponds to the time direction) with $\rho = 1, 2$ or $3$ and a
single non-zero component of momentum, $|p_\rho| = 2\pi/L$. The second
moment is extracted from $R^P_{\{\rho\mu4\};4}(t,\vec{p})$ with at
least two non-zero components of momentum. We take $\rho, \mu = 1, 2$
or $3$ with $\rho \ne \mu$ and $|p_\rho| = |p_\mu| = 2\pi/L$. We
present more details in Sec.~\ref{subsec:results}.

Apart from isolating the moments of the DAs as much as possible by
cancelling $Z_P$, $f^{\textrm{bare}}_P$ and most of the energy
dependence from Eqs.~\eqref{eq:C5rm} and \eqref{eq:C5rmn}, the ratios
also simplify the effect of mixing with total derivative operators.
These operators have matrix elements proportional to \eqref{eq:C5r}
with which we build a ratio similar to \eqref{eq:PSMratio}. Hence the
contribution of the mixing term becomes trivial and does not have to
be computed explicitly. It enters as an additive constant when
renormalising the bare moments as we will discuss later.

\subsection{$\boldsymbol{\VFM_V}$ and $\boldsymbol{\VSM_V}$ from Correlation
  Function Ratios}

The treatment of the vector meson's longitudinal distribution
amplitude is analogous. We consider the following two-point
correlation functions:
\begin{subequations}
\begin{align}
  C_{V_\mu V_\nu}(t,\vec{p}) &\equiv \sum_{\vec{x}} e^{i \vec{p}\cdot\vec{x}}
  \bra{0} V_\mu(t,\vec{x}) V_\nu^\dagger(0) \ket{0}, \\
  C_{\{\rho\mu\}\nu}(t,\vec{p}) &\equiv \sum_{\vec{x}} e^{i
    \vec{p}\cdot\vec{x}} \bra{0} \op{O}_{\{\rho\mu\}}(t,\vec{x}) V_\nu^\dagger(0)
  \ket{0}, \\
  C_{\{\rho\mu\nu\}\sigma}(t,\vec{p}) &\equiv \sum_{\vec{x}} e^{i
    \vec{p}\cdot\vec{x}} \bra{0} \op{O}_{\{\rho\mu\nu\}}(t,\vec{x})
  V_\sigma^\dagger(0) \ket{0}.
\end{align}
\end{subequations}
Again, the source operators $V^\dagger(0)$ are smeared. We define the
bare longitudinal decay constant of a vector meson $V$, with
polarisation index $\lambda$ and polarisation vector
$\varepsilon_\mu^{(\lambda)}$, by $\bra{0}V_\mu\ket{V(p,\lambda)}
\equiv f^{\textrm{bare}}_Vm_V\varepsilon_\mu^{(\lambda)}$. Then, at
large Euclidean times $t$ and $T - t$, the correlation functions
defined above may be written:
\begin{widetext}
\begin{subequations}
\begin{align}
  C_{V_\mu V_\nu}(t,\vec{p}) &\rightarrow
 %
  \frac{-(f_V^\mathrm{bare}m_V)^2e^{-E_VT/2}\cosh((t-T/2)E_V)}{E_V}
  \left(-g_{\mu\nu}+\frac{p_\mu p_\nu}{m_V^2} \right), \\
  C_{\{\rho\mu\}\nu}(t,\vec{p}) &\rightarrow
  \frac{-i(f_V^\mathrm{bare}m_V)^2e^{-E_VT/2}\VFMb\sinh((t-T/2)E_V)}{E_V}
    \,\frac{1}{2} \left(-g_{\rho\nu}p_\mu -g_{\mu\nu}p_\rho +
    \frac{2p_\rho p_\mu p_\nu}{m_V^2} \right), \\
  C_{\{\rho\mu\nu\}\sigma}(t,\vec{p}) &\rightarrow
%
  \frac{(f_V^\mathrm{bare}m_V)^2e^{-E_VT/2}\VSMb\sinh((t-T/2)E_V)}{E_V}
   \,\frac{1}{3} \left(-g_{\rho\sigma}p_\mu p_\nu
   -g_{\mu\sigma}p_\rho p_\nu -g_{\nu\sigma}p_\rho p_\mu + \frac{3p_\rho p_\mu
   p_\nu p_\sigma}{m_V^2} \right),
\end{align}
\end{subequations}
where we have used the completeness relation for the polarisation
vectors of massive vector particles,
$\sum_\lambda\varepsilon_\mu^{(\lambda)}\varepsilon_\nu^{*(\lambda)} =
-g_{\mu\nu} + p_\mu p_\nu/m_V^2$. We extract bare values for the first
and second moments from the following ratios:
\begin{subequations}
\label{eq:Vratio}
\begin{align}
  \label{eq:VFMratio}
  R^V_{\{\rho\mu\}\nu}(t,\vec{p}) &\equiv
  \frac{C_{\{\rho\mu\}\nu}(t,\vec{p})}{\frac{1}{3}\sum_i
    C_{V_iV_i}(t,\vec{p}=0\;)}
    \rightarrow
    -i\VFMb\tanh((t-T/2)E_V) \,\frac{1}{2} \left(-g_{\rho\nu}p_\mu
    -g_{\mu\nu}p_\rho + \frac{2p_\rho p_\mu p_\nu}{m_V^2} \right), \\
  \label{eq:VSMratio}
  R^V_{\{\rho\mu\nu\}\sigma}(t,\vec{p}) &\equiv
  \frac{C_{\{\rho\mu\nu\}\sigma}(t,\vec{p})}{\frac{1}{3}\sum_i
    C_{V_iV_i}(t,p_i=0,|\vec{p}|=\tpoL\;)}\nonumber\\
  &\rightarrow \VSMb\tanh((t-T/2)E_V)
    \,\frac{1}{3} \left(-g_{\rho\sigma}p_\mu p_\nu
    -g_{\mu\sigma}p_\rho p_\nu -g_{\nu\sigma}p_\rho p_\mu + \frac{3p_\rho p_\mu
      p_\nu p_\sigma}{m_V^2} \right),
\end{align}
\end{subequations}
\end{widetext}
where the index $i$ runs over spatial dimensions only. We obtain the
first moment from $R^V_{\{\rho4\}\nu}(t,\vec{p})$ at $\vec{p} = 0$ by
  taking $\rho = \nu = 1, 2$ or 3. The second moment is obtained from
  $R^V_{\{\rho\mu\nu\}\sigma}(t,\vec{p})$ by taking, for example, $\nu
  = 4, \rho = 1, \mu = \sigma = 2$ and a single non-zero component of
  $\vec{p}$ in the 1-direction.

\section{Numerical Simulations and Results}
\label{sec:numerics}

\subsection{Simulation Details}
\label{subsec:simdet}

Our numerical calculations are based upon gauge field configurations
drawn from the joint datasets used for the broader UKQCD/RBC
domain-wall fermion phenomenology programme. Configurations were
generated with $N_f=2+1$ flavours of dynamical domain-wall fermions
and with the Iwasaki gauge action, using the Rational Hybrid Monte
Carlo (RHMC)~\cite{Clark:2006fx} algorithm on QCDOC
computers~\cite{Boyle:2003ue,Boyle:2003mj,Boyle:2005gf} running the
Columbia Physics System (CPS) software~\cite{cps} and the
BAGEL~\cite{boyle-bagel-cpc,Bagel} assembler generator.

Our set of gauge configurations includes data with two different
volumes but at a single lattice spacing, thus giving us some
indication of the size of finite volume effects but no ability to
perform a continuum extrapolation. We therefore have an unavoidable
systematic error which is, however, formally of
$O(a^2\Lambda^2_\mathrm{QCD})$ due to the automatic $O(a)$-improvement
of the DWF action and operators. In the future, this limitation will
be overcome by performing the analysis with a dataset with a finer
lattice spacing with the same action (such a dataset is now available
and is currently being calibrated). In the meantime, following the
UKQCD/RBC procedure for these configurations~\cite{Allton:2008pn}, we
ascribe a $4\%$ uncertainty as the discretisation error on the
moments. For both volumes, we have a single dynamical strange quark
mass, close to its physical value. We use several independent
ensembles with different light-quark masses ($m_u=m_d$), all heavier
than those found in nature. The hadronic spectrum and other properties
of these configurations have been studied in detail and the results
have been presented in \cite{Allton:2007hx} (for the lattice volume
$(L/a)^3 \times T/a = 16^3 \times 32$) and \cite{Allton:2008pn} (for
the lattice volume $24^3 \times 64$). In both cases the length of the
fifth dimension is $L_s = 16$.

The choice of bare parameters in our simulations is $\beta = 2.13$ for
the bare gauge coupling, $am_s = 0.04$ for the strange quark mass and
$am_q = 0.03,\;0.02,\;0.01$ and, in the $24^3$ case only, $0.005$ for
the bare light-quark masses. A posteriori, the strange quark mass is
found to be about 15\% larger than its physical value. The lattice
spacing is found to be $a^{-1} = 1.729(28)\gev$~\cite{Allton:2008pn},
giving physical volumes of $(1.83\;\mathrm{fm})^3$ and
$(2.74\;\mathrm{fm})^3$. The lattice spacing and physical quark masses
were obtained using the masses of the $\pi$ and $K$ pseudoscalar
mesons and the triply-strange $\Omega$ baryon. The quark masses
obtained in the $24^3$ study are shown in Table~\ref{tab:24qmasses}.
Owing to the remnant chiral symmetry breaking, the quark mass has to
be corrected additively by the residual mass in the chiral limit,
$a\mres = 0.00315(2)$~\cite{Allton:2008pn}. The physical pion masses
are as follows:
\begin{equation}
m_\pi \simeq
\begin{cases}
  670\mev & am_q=0.03\\
  555\mev & am_q=0.02\\
  415\mev & am_q=0.01\\
  330\mev & am_q=0.005
\end{cases}
\end{equation}

\begin{table*}
  \caption{\label{tab:24qmasses}Lattice scale and unrenormalised quark
    masses in lattice units, from the $24^4$
    lattices~\cite{Allton:2008pn}. Note $\tilde{m}_X \equiv m_X +
    \mres$. Only the statistical errors are given here.}
  \begin{ruledtabular}
    \begin{tabular}{c|cccccc}
      $a^{-1}$ [\Gev] & $a$ [{\rm fm}] & $am_{ud}$ &
      $a\widetilde{m}_{ud}$ & $am_s$ & $a\widetilde{m}_s$ &
      $a\widetilde{m}_{ud}:a\widetilde{m}_s$ \\\hline
     $1.729(28)$ & $0.1141(18)$ & $-0.001847(58)$ & $0.001300(58)$ &
     $0.0343(16)$ & $0.0375(16)$ & 1:28.8(4)\\
    \end{tabular}
  \end{ruledtabular}
\end{table*}
\begin{table}
  \caption{\label{tab:16datasets}Parameters for our $16^3$ dataset,
    which corresponds largely to that of~\cite{Allton:2007hx}. The
    range and measurement separation $\Delta$ are specified in
    molecular dynamics time units. $N_\mathrm{meas}$ is the number of
    measurements for each source position $t_\mathrm{src}$. The total
    number of measurements is therefore $N_\mathrm{meas}\times
    N_\mathrm{src}$, where $N_\mathrm{src}$ is the number of different
    values for $t_\mathrm{src}$. In the right-most column, XY-XY
    denotes contraction of two quark propagators with X-type smearing
    at source and Y-type smearing at sink: G = Gaussian wavefunction,
    L = point.}
  \begin{ruledtabular}
    \begin{tabular}{c|cccccc}
      $m_l$ & Range & $\Delta$ & $N_\mathrm{meas}$ &
      $t_\mathrm{src}$ locations & Smearing\\
      \hline
      0.01 & 500--3990 & 10 & 350 & 0,\,8,\,16,\,24 & GL-GL \\
      0.02 & 500--3990 & 10 & 350 & 0,\,8,\,16,\,24 & GL-GL \\
      0.03 & 4030--7600 & 10 & 358 & 0,\,16 & GL-GL \\
    \end{tabular}
  \end{ruledtabular}
\end{table}
\begin{table}
  \caption{\label{tab:24datasets}Parameters for our $24^3$ dataset,
    which corresponds to the unitary part of the dataset
    of~\cite{Allton:2008pn}. Columns as in Table~\ref{tab:16datasets}
    with addition of H = gauge-fixed hydrogen S-wave smearing.}
  \begin{ruledtabular}
    \begin{tabular}{c|cccccc}
      $m_l$ & Range & $\Delta$ & $N_\mathrm{meas}$ &
      $t_\mathrm{src}$ locations & Smearing\\ \hline
      0.005 & 900--4480 & 20 & 180 & 0,\,32,\,16 & HL-HL\\
      0.01 & 800--3940 & 10 & 315 & 0,\,32 & GL-GL \\
      0.02 & 1800--3580 & 20 & 90 & 0,\,32 & HL-HL \\
      0.03 & 1260--3040 & 20 & 90 & 0,\,32 & HL-HL \\
    \end{tabular}
  \end{ruledtabular}
\end{table}

Measurements were performed using the UKhadron software package that
makes use of both the BAGEL DWF inverter~\cite{boyle-bagel-cpc,Bagel}
and elements of the SciDAC software library stack including the Chroma
LQCD library~\cite{Edwards:2004sx} and QDP++. The details are
summarised in Tables~\ref{tab:16datasets} and \ref{tab:24datasets}. We
restrict our analysis to the unitary data for which the valence and
sea quark masses are the same (partially-quenched data was used
extensively in the studies of the chiral behaviour of the spectrum and
decay constants in~\cite{Allton:2008pn}). On the $16^3$ lattice, our
dataset differs from that used in~\cite{Allton:2007hx} in that the
Markov chains have been extended for the heaviest light quark mass to
give additional statistics, using an improved algorithm that
decorrelated topology rather more quickly.

In order to improve the statistical sampling of the correlation
functions, on each configuration we have averaged the results obtained
from either $2$, $3$ or $4$ sources spaced out along a lattice
diagonal. In the $16^3$ case, for example, the sources used are at the
origin, at $(4,4,4,8)$, $(8,8,8,16)$ and $(12,12,12,24)$. Statistical
errors for observables are estimated using single-elimination
jackknife, with measurements made on the same configuration but at
different source positions put in the same jackknife bin because of
the correlations expected between them. In order to lessen the effect
of autocorrelations, we follow the same blocking procedures as
in~\cite{Allton:2007hx} and~\cite{Allton:2008pn}. In the $16^3$ case,
the span of the measurements in each block covers $50$ molecular
dynamics time units. In the $24^3$ case, for the $m_qa = 0.005$ and
$am_q = 0.01$ ensembles, each jackknife bin contains measurements from
every $80$ molecular dynamics time units, while for the $am_q = 0.02$
and $am_q = 0.03$ ensembles each bin contains measurements from every
$40$ molecular dynamics time units in order to have a reasonable
number of bins for the analysis.

We use source smearing to improve the overlap with the mesonic states,
either gauge-fixed hydrogen $S$-wavefunction
smearing~\cite{Boyle:1999gx} with radius $r = 3.5$ in lattice units or
gauge invariant Gaussian smearing~\cite{Allton:1993wc} with radius $r
= 4$.

\subsection{Results}
\label{subsec:results}

\begin{figure*}
  \includegraphics[width=0.48\textwidth,
                   bb=90 420 436 658]{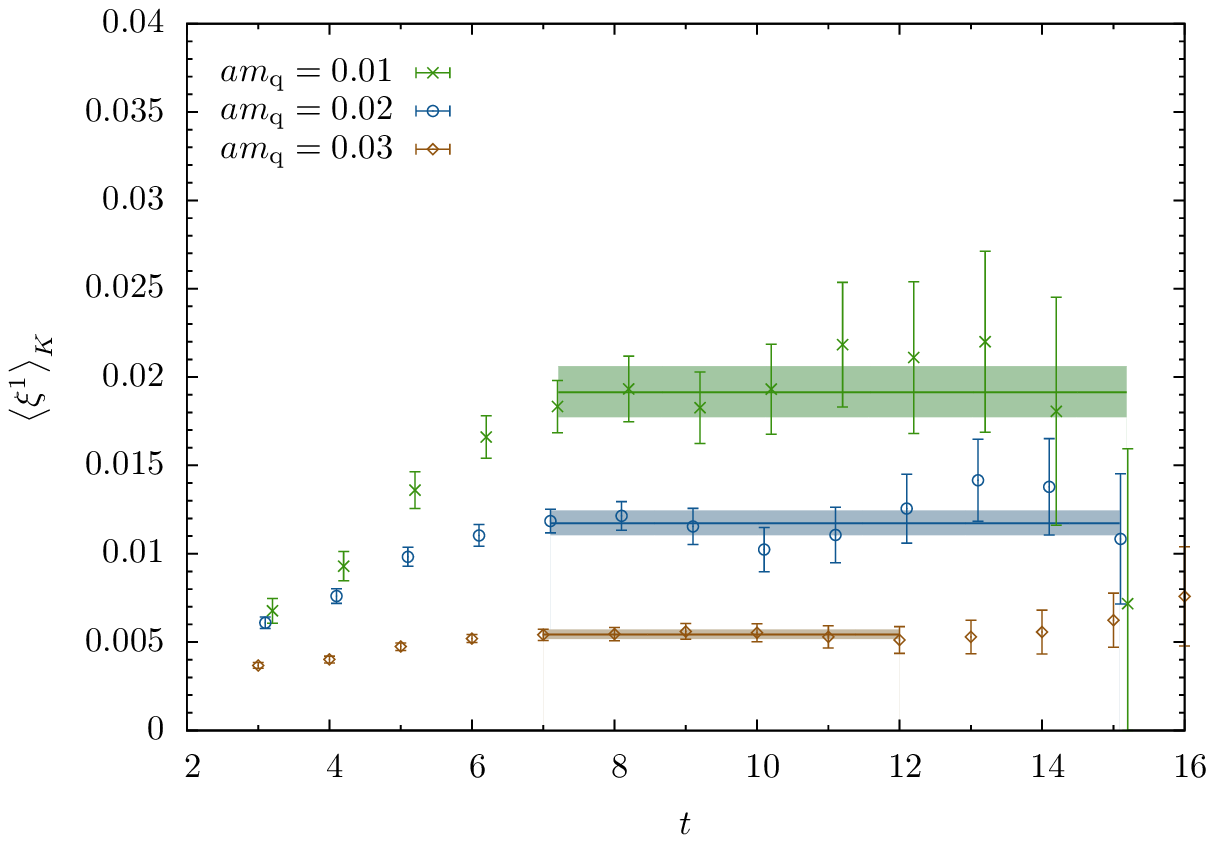}
  \hfill
  \includegraphics[width=0.48\textwidth,
                   bb=90 420 436 658]{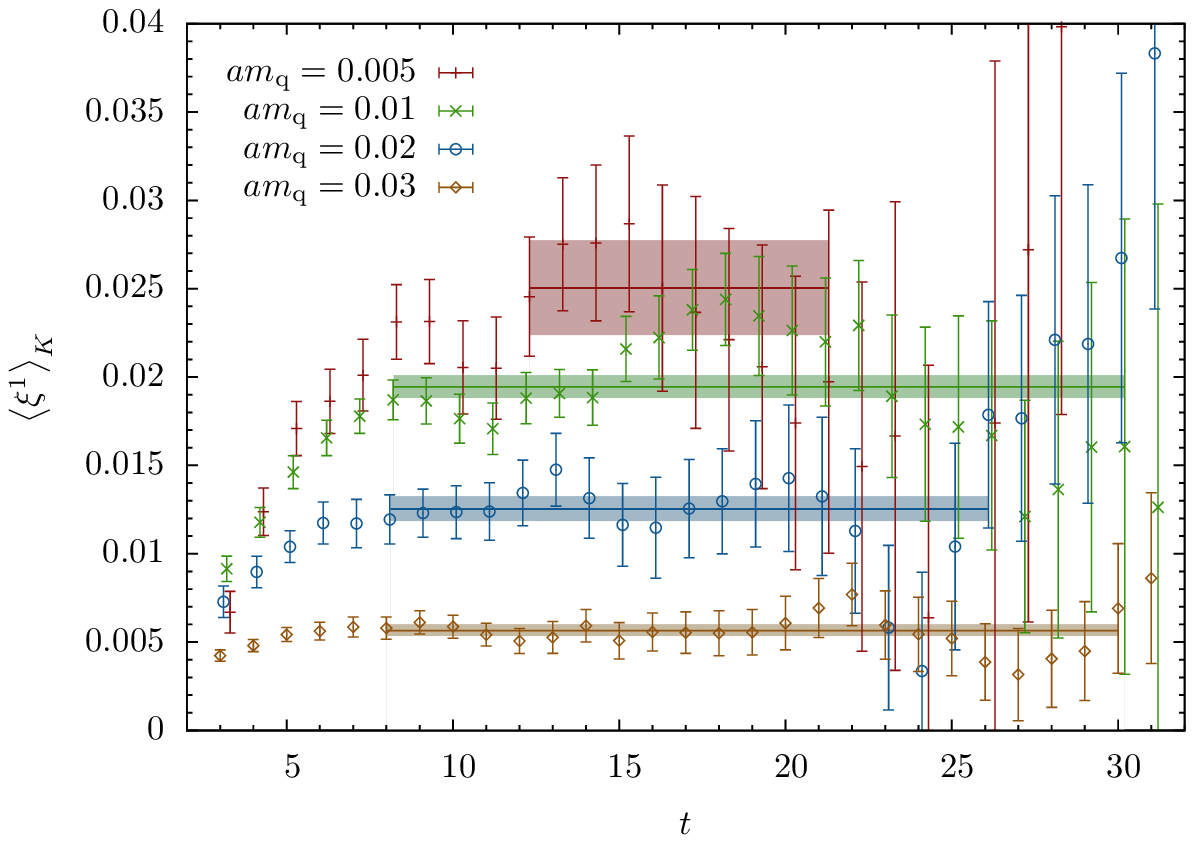}
  \caption{\label{fig:corrsK1}Results for $\PFM_K^\mathrm{bare}$ as a
    function of the time, on the $16^3$ (left) and $24^3$ (right)
    lattices. The shaded band shows the fit range, fitted value and
    its error.}
\end{figure*}
In order to extract $\PFM_K$ from the ratio
$R^P_{\{\rho\mu\};\nu}(t,\vec{p})$ defined in \eqref{eq:PFMratio}, we
need the two correlation functions to be measured at $|\vec{p}| \ne
0$. Since we expect hadronic observables with larger lattice momenta
to have larger lattice artefacts and statistical errors, we restrict
the choice of indices to $\rho=\nu=4$ and $\mu=1,2$ or 3 with $|{\vec
  p}\,|=2\pi/L$ (i.e., $p_\mu=\pm 2\pi/L$ with the remaining two
components of ${\vec p}$ equal to 0). $\PFM_K^\mathrm{bare}$ can then
be obtained from the ratio at large times:
\begin{equation}
    R^P_{\{4 k\};\,4}(t,p_k=\pm 2\pi/L)=
       \pm\, i\,\tpoL\,\PFM^\mathrm{bare}\,,
\end{equation}
with $|\vec{p}|=2\pi/L$ and $k=1,2,3$. The plots in
Fig.~\ref{fig:corrsK1} show our results for $\PFM_K^\mathrm{bare}$ as
a function of $t$ obtained from the ratio $R^P_{\{4
  k\};\,4}(t,p_k{=}\pm 2\pi/L)$ for the four values of the light-quark
mass, combining results at $t$ with those at $T-t-1$. The results have
been averaged over the three values of $k$ and, in total, the $6$
equivalent lattice momenta with $|\vec{p}|=2\pi/L$.

To obtain $\PSM_{\pi,K}^\mathrm{bare}$ from the ratio
$R^P_{\{\rho\mu\nu\};\sigma}(t,\vec{p})$ defined in
\eqref{eq:PSMratio} we need two non-zero components of momentum, so we
use
\begin{multline}
  R^P_{\{4 j k\};\,4}(t,p_j=\pm 2\pi/L, p_k = \pm 2\pi/L)=\\
    -(\pm\tpoL)(\pm\tpoL)\PSM^\mathrm{bare}\,,
\end{multline}
with $|\vec{p}|=\sqrt{2}\,2\pi/L$, $k,j=1,2,3$ and $k \ne j$. We average
over all $4$ momentum combinations appropriate to each of the $3$
possible Lorentz index choices.

\begin{figure*}
  \includegraphics[width=0.48\textwidth,
                   bb=90 420 436 658]{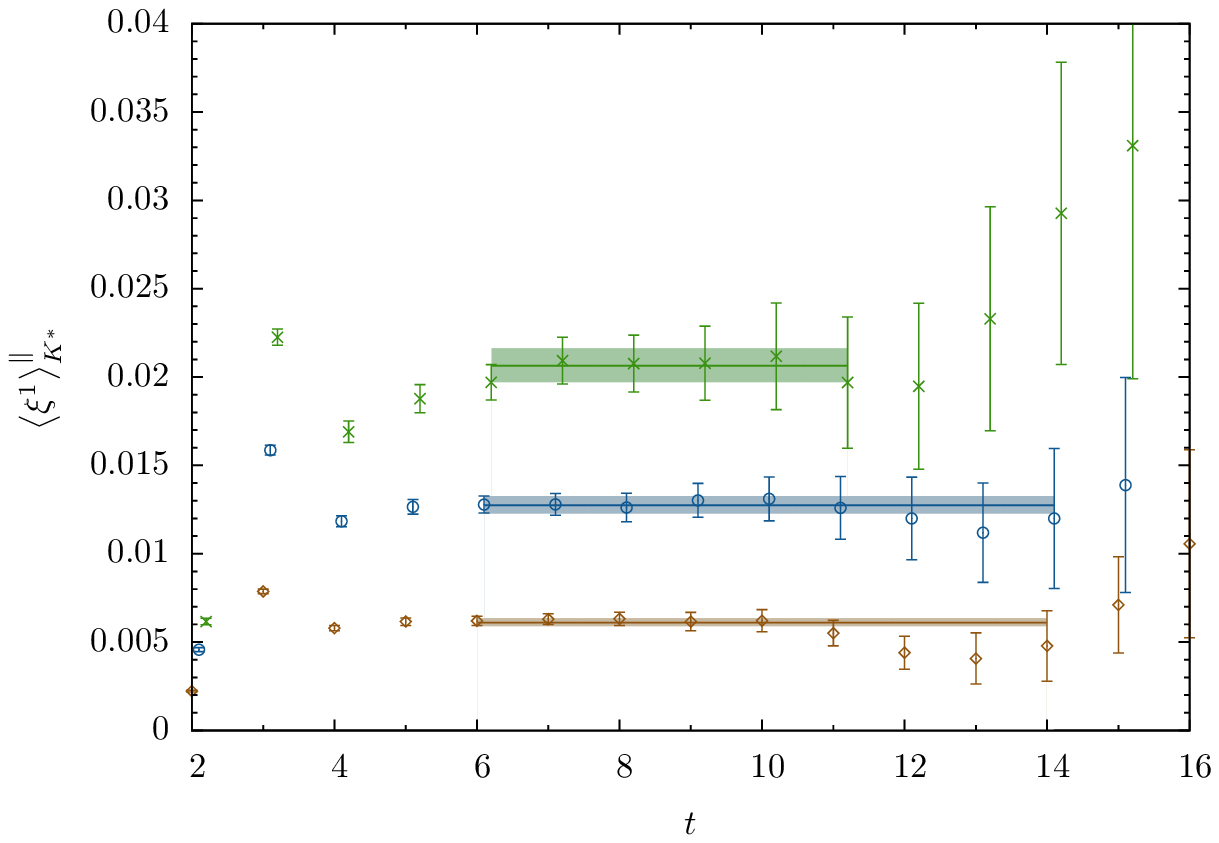}
  \hfill
  \includegraphics[width=0.48\textwidth,
                   bb=90 420 436 658]{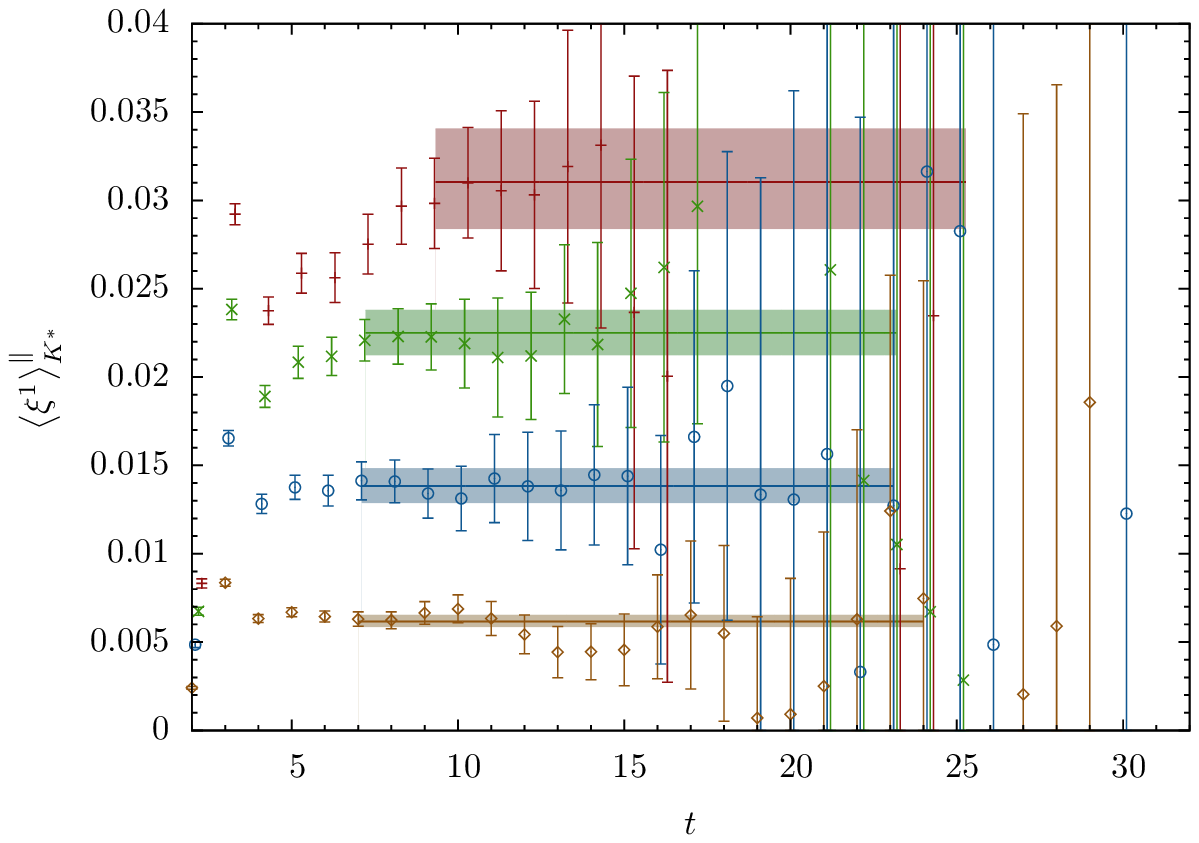}
  \caption{\label{fig:corrsV1}Results for $\VFMb_{K^*}$ as a function of the
    time, on the $16^3$ (left) and $24^3$ (right) lattices. Symbols as in
    Fig.~\ref{fig:corrsK1}}
\end{figure*}

We may extract $\VFMb_{K^*}$ from the ratio
$R^V_{\{\rho\mu\}\nu}(t,\vec{p})$ defined in \eqref{eq:VFMratio} by
considering only zero-momentum correlation functions. In the
denominator, we average $C_{V_iV_i}(t,\vec{p}=0)$ over all $3$ spatial
directions. In the numerator, we average over
$C_{\{41\}1}(t,\vec{p}=0)$, $C_{\{42\}2}(t,\vec{p}=0)$ and
$C_{\{43\}3}(t,\vec{p}=0)$. Results are shown in
Fig.~\ref{fig:corrsV1}.
$\VSMb_{K^*,\rho,\phi}$ is extracted from the ratio defined in
\eqref{eq:VSMratio} by averaging
$C_{V_iV_i}(t,p_i=0,|\vec{p}|=\tpoL\;)$ over all $4$ appropriate
momenta for all $3$ spatial directions in the denominator. In the
numerator we average over all possible combinations of
$C_{\{4ij\}i}(t,p_j = \pm \tpoL,|\vec{p}|=\tpoL)$ with $i \ne j$. In
principle we should include disconnected contributions in the $\phi$
correlation functions. We argue that these contributions are
Zweig-suppressed however and can therefore be neglected.

If picking the fit range was not straightforward, we considered the
correlation functions in the numerator and denominator separately. We
identified and excluded from our fits the region where the excited
states still contributed. We then chose the fit range aiming for a good
$\chi^2/\text{d.o.f.}$ and a stable fit with respect to small
variations of the lower bound of the range. Owing to the increasing
noise when $t$ gets larger, the fits are insensitive to the upper
bound of the fit range. 

The $16^3$ and $24^3$ bare results are given in
Tables~\ref{tab:16results} and \ref{tab:24results} respectively,
complete with linear chiral extrapolations which, as we shall discuss
in the next section, can be justified using chiral perturbation theory
(at least in the pseudoscalar case).

\subsection{Quark Mass Extrapolations}
\label{subsec:chiral}

\begin{figure*}
 \includegraphics[width=0.48\hsize,bb=169 418 497 660]{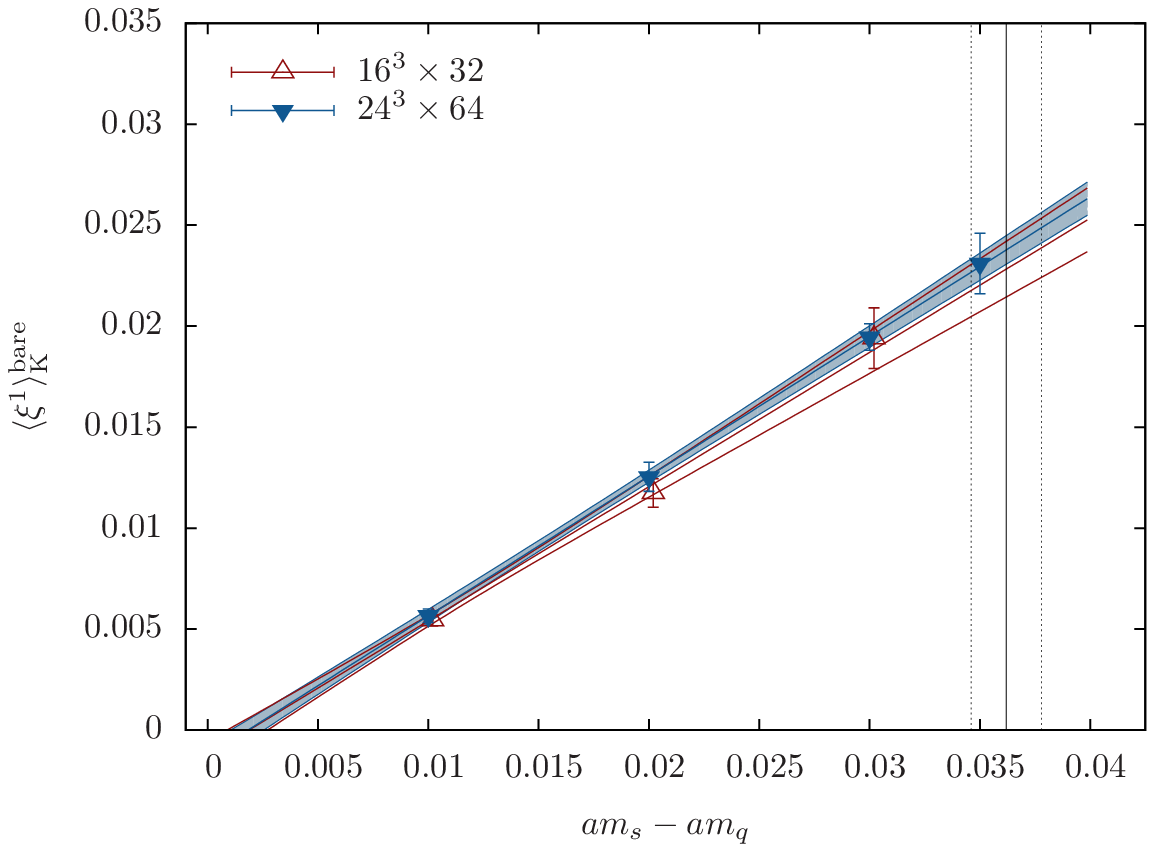}
 \hfill
 \includegraphics[width=0.48\hsize,bb=169 418 497 660]{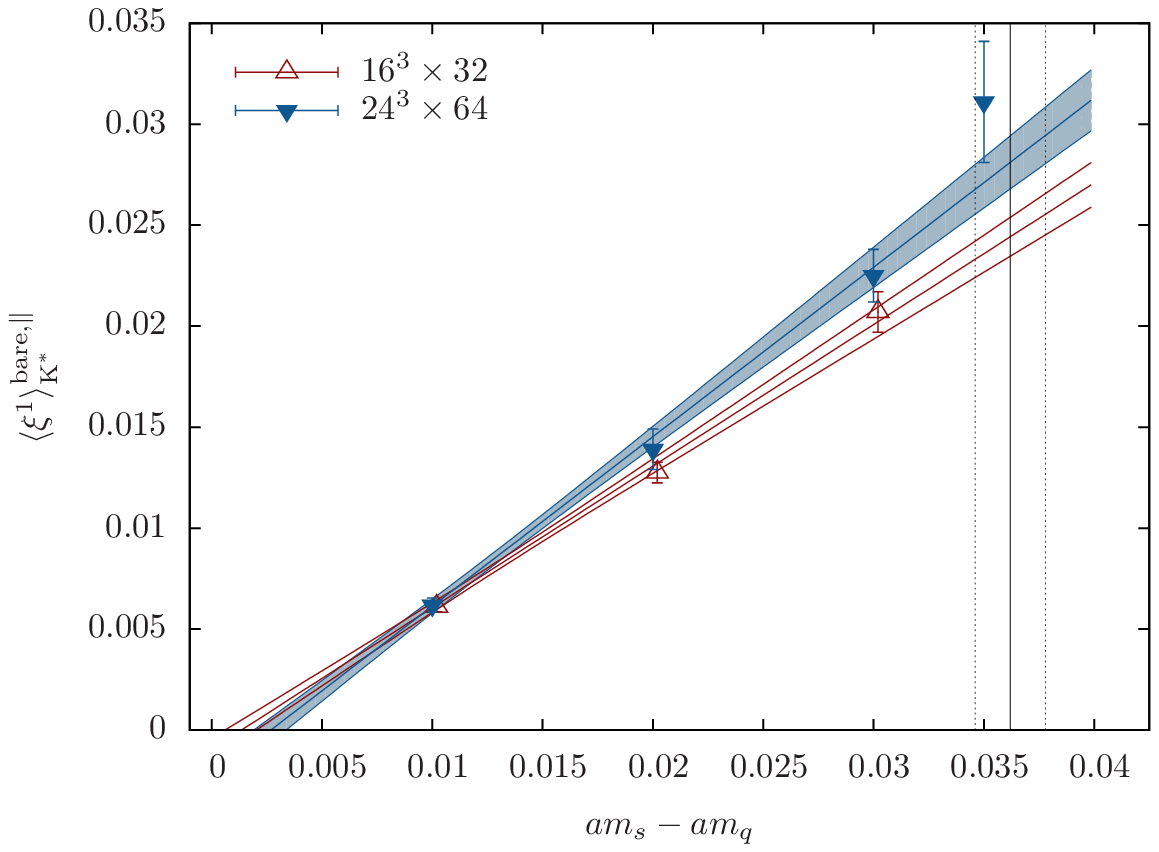}
  \caption{\label{fig:Kaon1st_chiral}Chiral extrapolations for
    $\PFM_K^\mathrm{bare}$ and $\VFMb_{K^*}$. The extrapolation to the
    physical point is shown by the vertical solid line, with
    uncertainty, dominated by the uncertainty in the physical strange
    mass, indicated by the dotted lines.}
\end{figure*}
In leading-order chiral perturbation theory~\cite{Chen:2003fp}, $\PFM_K$ is
proportional to $m_s - m_{u/d}$ without chiral logarithms:
\begin{equation}
  \langle\xi^1\rangle_K = \frac{8B_0}{f^2}(m_s - m_{u/d})b_{1,2}\;,
\end{equation}
where $f$ and $B_0$ denote the usual chiral perturbation theory
parameters and $b_{1,2}$ is a Wilson coefficient introduced
in~\cite{Chen:2003fp}. Our data shows clearly the effects of SU(3)
symmetry breaking and is compatible with this expectation. We
therefore perform a linear extrapolation in $a(m_s - m_q)$ to the
physical point $a(m_s - m_{ud})$, as shown in
Fig.~\ref{fig:Kaon1st_chiral}. The second error quoted in the results
in the chiral-limit for the first moments in
Tables~\ref{tab:16results} and~\ref{tab:24results} is due to the
uncertainty in this physical point (determined using the quark masses
in Table~\ref{tab:24qmasses}). In this way we deal simultaneously with
the usual light-quark mass extrapolation and with the strange quark
mass extrapolation which is necessitated by our strange quark mass
being approximately $15\%$ too heavy. We have not constrained our fit
to vanish in the SU(3) limit.

A similar linear behaviour is seen for $\VFMb_{K^*}$ (see
Fig.~\ref{fig:Kaon1st_chiral}), so we follow the same extrapolation
procedure. We note a hint of a finite volume effect in the $K^*$ case
but not in the $K$ case, which is contrary to what we would expect.
Where we have $K^*$ results for both volumes at the same
light-quark mass, however, they agree within the statistical
uncertainties.

\begin{figure*}
\includegraphics[width=0.48\hsize,bb=108 417 435 657]{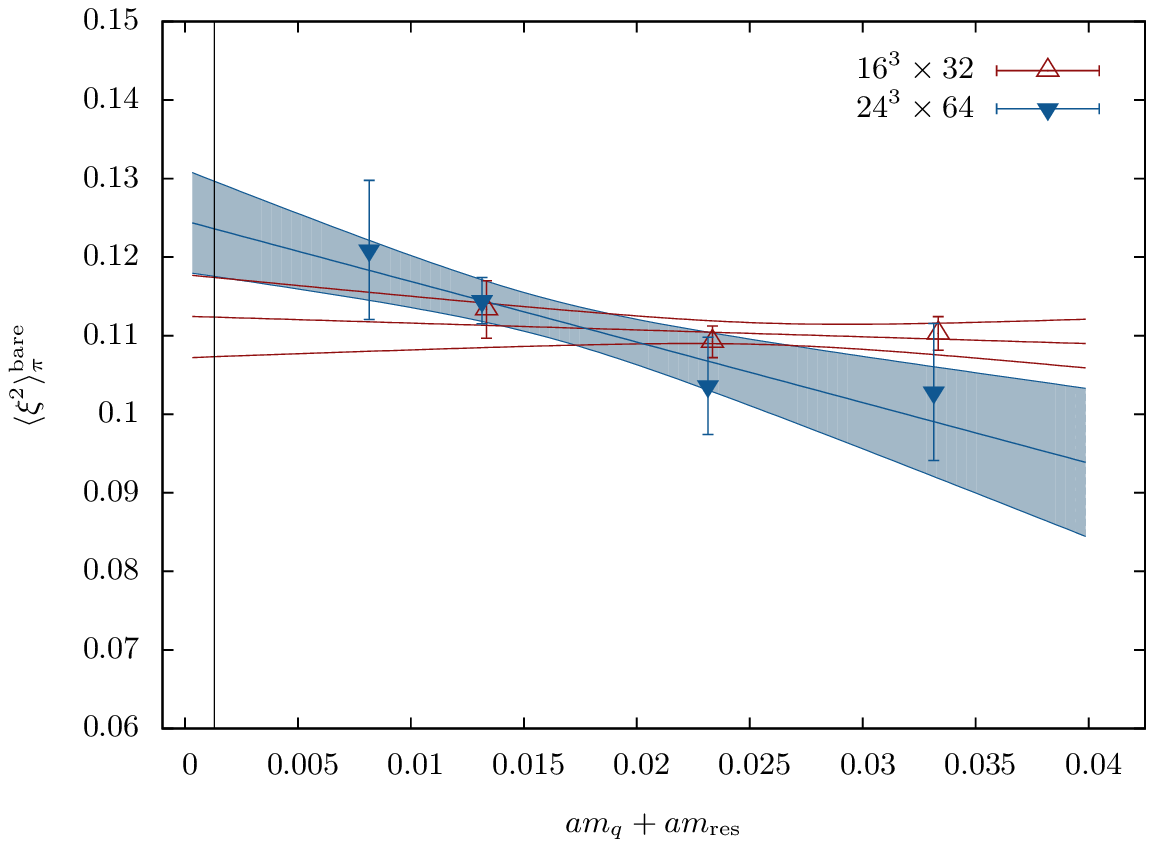}
\hfill
\includegraphics[width=0.48\hsize,bb=108 417 435 657]{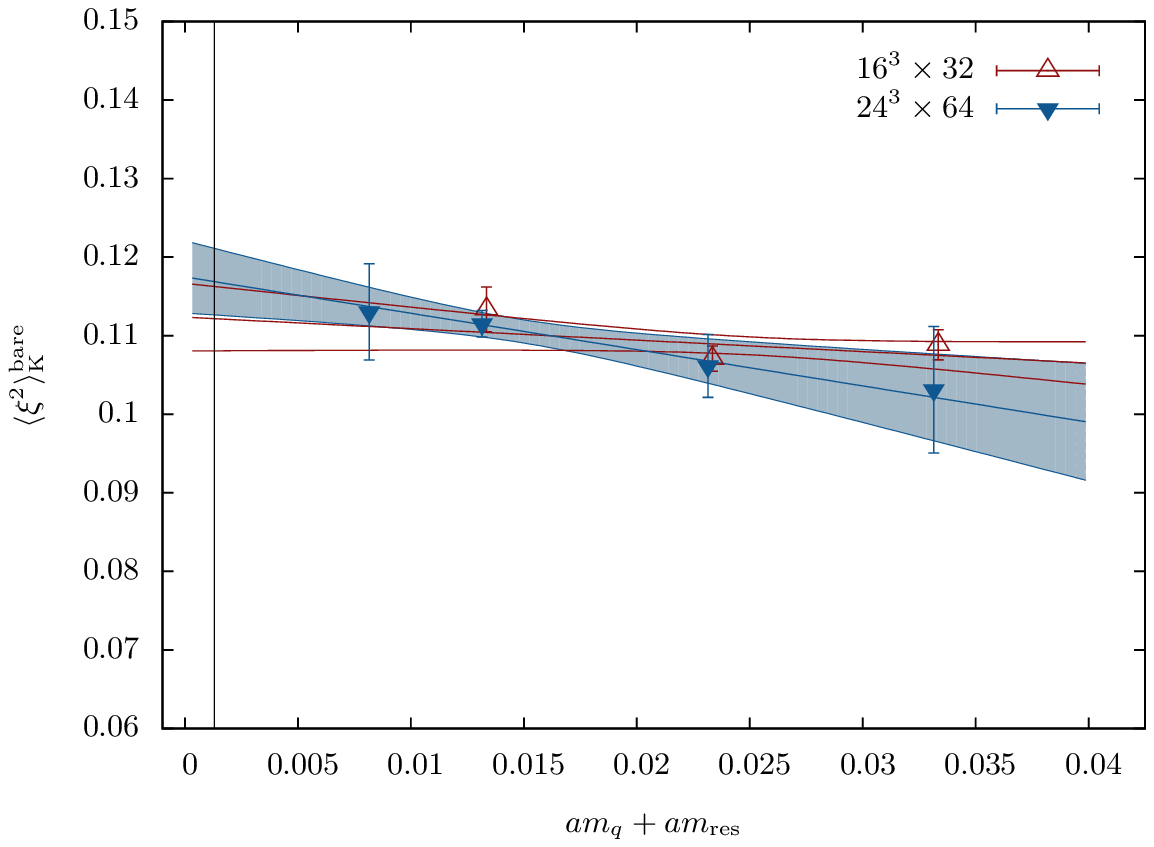}\\[1ex]
\includegraphics[width=0.48\hsize,bb=108 417 435 657]{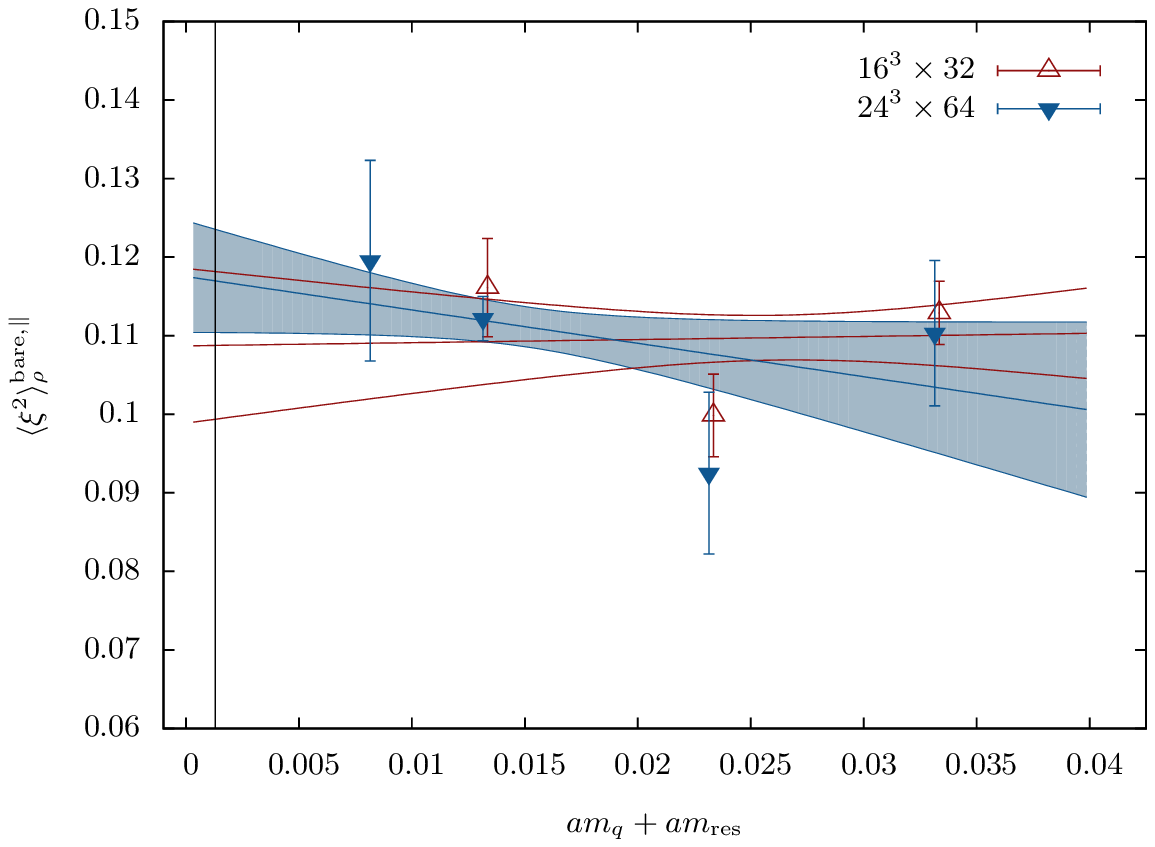}
\hfill
\includegraphics[width=0.48\hsize,bb=108 417 435 657]{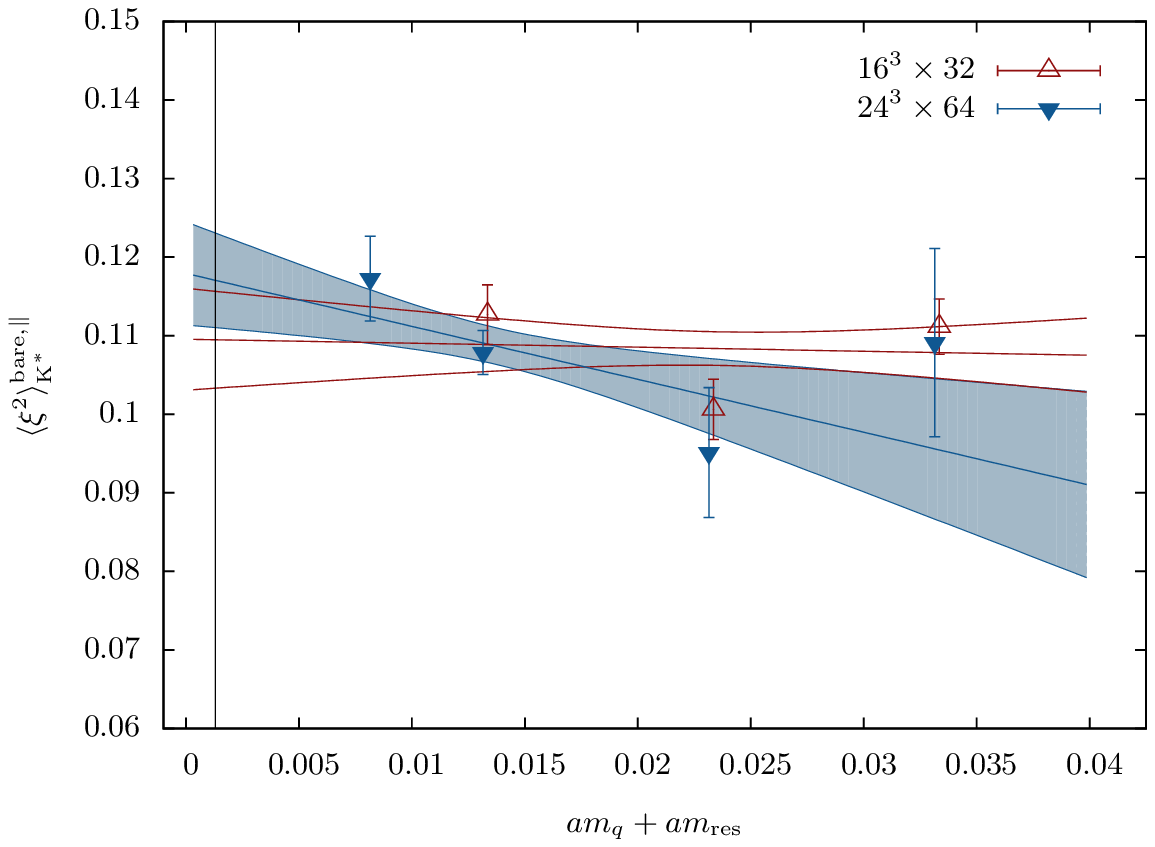}\\[1ex]
\includegraphics[width=0.48\hsize,bb=108 417 435 657]{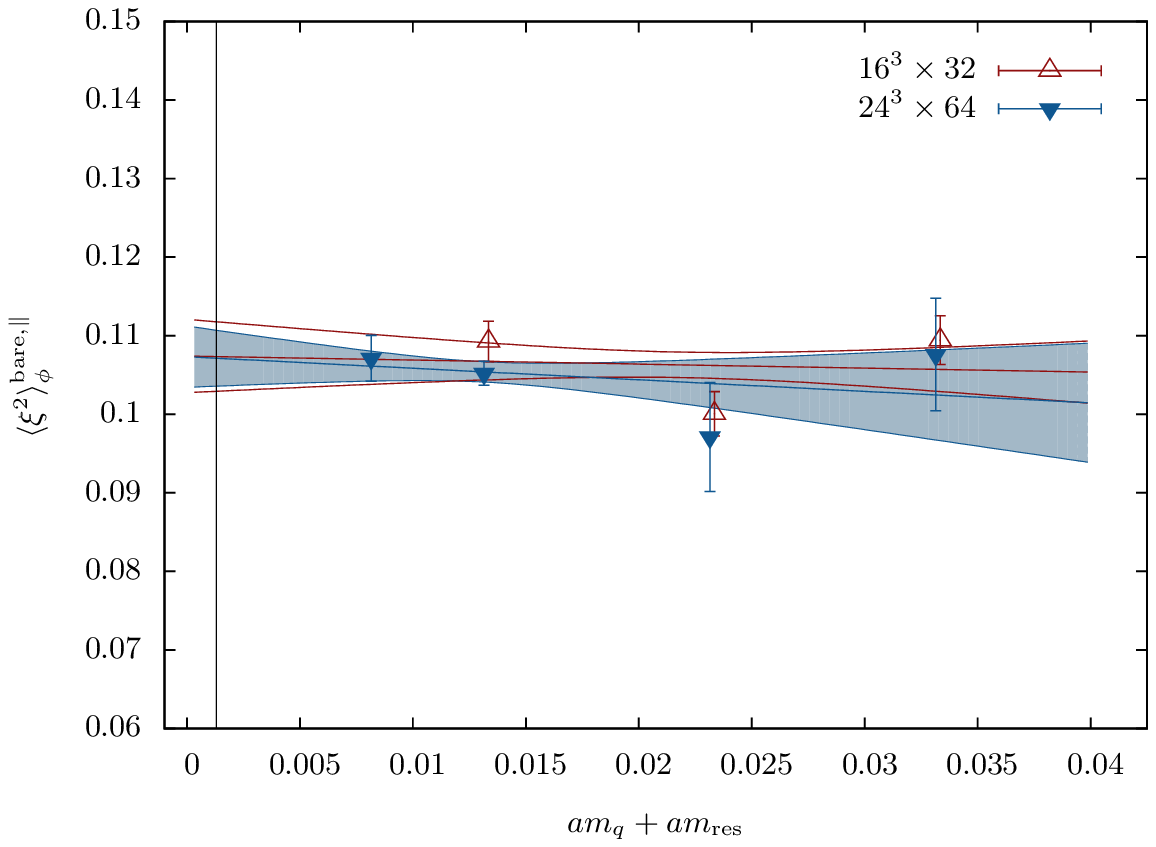}
\caption{Chiral extrapolations for $\PSM_{\pi}^\mathrm{bare}$,
  $\PSM_{K}^\mathrm{bare}$, $\VSMb_{\rho}$, $\VSMb_{K^*}$ and
  $\VSMb_{\phi}$. The physical value for $am_q + a\mres$ is
  shown by the solid vertical line in each case.\label{fig:a-z}}
\end{figure*}

For the second moments, we also have some guidance from chiral
perturbation theory~\cite{Chen:2005js}; there is no non-analytic
dependence at 1~loop and we should fit linearly in $m^2_\pi$. The
dependence on the quark masses is very mild in any case and in fact
our results for the $\rho$, $K^*$ and $\phi$ agree within the
statistical errors. Therefore we perform a linear extrapolation in the
light quark masses and neglect the effect of the too-heavy strange
quark mass (see Fig.~\ref{fig:a-z}). We see no indication for finite
size effects in the second moments when we compare the data points on
the two different lattice volumes. They agree within their statistical
errors.
\begin{table*}
  \caption{\label{tab:16results}Summary of results for the bare values
    of the distribution amplitude moments on the $16^3$ lattices. The
    chiral extrapolations are discussed in Sec.~\ref{subsec:chiral}, and
    the errors are statistical and (in the first moment case) due to the
    uncertainty in the physical point for the chiral extrapolation.}
  \begin{ruledtabular}
    \begin{tabular}{l|llllll}
     $am_{ud}$ & 0.03  & 0.02   & 0.01   & 0.005 & $\chi$-limit \\
     \hline
     $\PSM_\pi^\mathrm{bare}$ & 0.110(2) & 0.109(2) & 0.113(4) & - & 0.112(5) \\
     $\PFM_K^\mathrm{bare}$ & 0.00543(27) & 0.01174(71) & 0.0194(15) & - & 0.0228(14)(11) \\
     $\PSM_K^\mathrm{bare}$ & 0.109(2) & 0.107(2) & 0.113(3) & - &
     0.112(4) \\
     $\VSMb_\rho$& 0.113(4) & 0.100(5) & 0.116(6) & - & 0.109(10) \\
     $\VFMb_{K^*}$ & 0.00610(24) & 0.01275(51) & 0.0207(10) & - &
     0.02443(96)(107) \\
     $\VSMb_{K^*}$ & 0.111(4) & 0.101(4) & 0.113(4) & - & 0.110(6) \\
     $\VSMb_\phi$& 0.109(3) & 0.100(3) & 0.109(3) & - & 0.107(5)   \\
    \end{tabular}
  \end{ruledtabular}
\end{table*}
\begin{table*}
  \caption{\label{tab:24results}Summary of results for the bare values
    of the distribution amplitude moments on the $24^3$ lattices}
  \begin{ruledtabular}
    \begin{tabular}{l|llllll}
      $am_{ud}$ & 0.03  & 0.02 & 0.01 & 0.005 & $\chi$-limit \\
      \hline
      $\PSM_\pi^\mathrm{bare}$ & 0.103(9)  & 0.104(6)  & 0.114(3)  & 0.121(9) & 0.125(7) \\
      $\PFM_K^\mathrm{bare}$   & 0.00566(33) & 0.01254(72) & 0.01946(65) & 0.0231(15) & 0.02377(71)(110) \\
      $\PSM_K^\mathrm{bare}$   & 0.103(8)  & 0.106(4)  & 0.112(2)  & 0.113(6) & 0.117(5) \\
      $\VSMb_\rho$& 0.110(9)  & 0.093(10) & 0.112(3)  & 0.120(13)& 0.118(7) \\
      $\VFMb_{K^*}$ & 0.00619(35) & 0.0139(10)& 0.0225(13)& 0.0311(30) & 0.0281(13)(14) \\
      $\VSMb_{K^*}$ & 0.109(12) & 0.095(8)  & 0.108(3)  & 0.117(5) & 0.118(7) \\
      $\VSMb_\phi$& 0.108(7)  & 0.097(7)  & 0.105(2)  & 0.107(3) & 0.107(4) \\
    \end{tabular}
  \end{ruledtabular}
\end{table*}

\section{Renormalisation of the Lattice Composite Operators}
\label{sec:renormalisation}

We now discuss the conversion of our bare lattice results to results
in the \MSbar\ scheme. To reduce systematic uncertainties we have
determined the renormalisation factors nonperturbatively in the
\ripm\ scheme, continuing the work in~\cite{Aoki:2007xm}, and convert
to \MSbar\ using $3$-loop continuum perturbation
theory~\cite{Gracey:2003mr, Gracey:2006zr}. We begin, however, with a
perturbative calculation of the renormalisation factors. The
perturbative results have been used previously
in~\cite{Donnellan:2007xr, Boyle:2008nj} and will provide a comparison
to the nonperturbative results. The contribution to the second moment
from mixing with a total-derivative operator is calculated
perturbatively only. We will see that this contribution is small and
is not accessible within the current nonperturbative scheme.

\subsection{Perturbative Renormalisation}
\label{sec:PR}

\begin{figure}
  \hbox to\hsize{\hss
    \includegraphics[width=0.8\hsize]{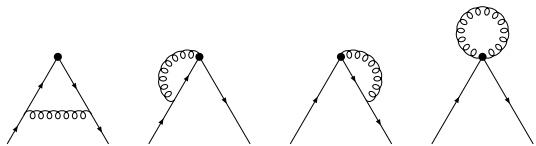}
    \hss}
  \caption{One-loop vertex diagrams evaluated in the perturbative renormalisation
    of the $1$st and $2$nd moment operators.}
  \label{fig:ptdiags}
\end{figure}

\begin{figure}
  \hbox to\hsize{\hss
    \includegraphics[width=0.4\hsize]{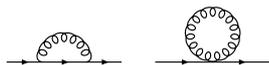}
    \hss}
  \caption{One-loop diagrams for the quarks' wavefunction renormalisation.}
  \label{fig:wavefnren}
\end{figure}

\begin{table*}
  \caption{Constants needed for the perturbative renormalisation of the
    first and second moment operators using domain-wall fermions and the
    Iwasaki gauge action ($c_1=-0.331$). $M$ is the domain-wall height,
    $c=\Sigma_1^\MSbar - \Sigma_1 + V^\MSbar - V$, $\cDD =
    \Sigma_1^\MSbar - \Sigma_1 + \VDD^\MSbar - \VDD$ and
    $\cdd=\Vdd^\MSbar-\Vdd$. $\Sigma_1$, $V$, $\VDD$ and $\Vdd$ are
    dependent on the gauge and the infrared regulator: Feynman gauge and
    a gluon mass are used here. $V$ was calculated
    in~\cite{Boyle:2006pw}, while $\VDD$ and $\Vdd$ have been calculated
    as part of this work.}
  \label{tab:VvsM}
  \begin{minipage}{.9\textwidth}
  \begin{ruledtabular}
    \begin{tabular}{D..1D..4D..4D..4D..4D..4D..4D..4}
      \multicolumn1c{$M$} &
      \multicolumn1c{$\Sigma_1$} &
      \multicolumn1c{$V$} &
      \multicolumn1c{$c$} &
      \multicolumn1c{$\VDD$} &
      \multicolumn1c{$\cDD$} &
      \multicolumn1c{$\Vdd$} &
      \multicolumn1c{$\cdd$} \\
      \hline
      0.1 & 4.6519 & -4.6297 & -0.9110 & -10.816 & 4.9838 & 0.5415 & 0.0279 \\
      0.2 & 4.5193 & -4.5614 & -0.8468 & -10.698 & 4.9982 & 0.4285 & 0.1409 \\
      0.3 & 4.4093 & -4.5101 & -0.7881 & -10.608 & 5.0179 & 0.3433 & 0.2262 \\
      0.4 & 4.3158 & -4.4678 & -0.7369 & -10.533 & 5.0362 & 0.2729 & 0.2966 \\
      0.5 & 4.2354 & -4.4311 & -0.6932 & -10.467 & 5.0509 & 0.2119 & 0.3575 \\
      0.6 & 4.1665 & -4.3980 & -0.6574 & -10.407 & 5.0603 & 0.1573 & 0.4122 \\
      0.7 & 4.1079 & -4.3673 & -0.6295 & -10.352 & 5.0639 & 0.1070 & 0.4625 \\
      0.8 & 4.0593 & -4.3381 & -0.6101 & -10.300 & 5.0604 & 0.0597 & 0.5098 \\
      0.9 & 4.0204 & -4.3097 & -0.5996 & -10.250 & 5.0489 & 0.0142 & 0.5552 \\
      1.  & 3.9915 & -4.2816 & -0.5988 & -10.200 & 5.0283 & -0.0303 & 0.5998 \\
      1.1 & 3.9731 & -4.2529 & -0.6090 & -10.151 & 4.9970 & -0.0749 & 0.6443 \\
      1.2 & 3.9664 & -4.2232 & -0.6321 & -10.100 & 4.9528 & -0.1205 & 0.6899 \\
      1.3 & 3.9727 & -4.1916 & -0.6700 & -10.047 & 4.8933 & -0.1682 & 0.7376 \\
      1.4 & 3.9943 & -4.1571 & -0.7261 & -9.9895 & 4.8147 & -0.2195 & 0.7889 \\
      1.5 & 4.0343 & -4.1182 & -0.8050 & -9.9267 & 4.7119 & -0.2764 & 0.8458 \\
      1.6 & 4.0974 & -4.0728 & -0.9135 & -9.8551 & 4.5771 & -0.3418 & 0.9112 \\
      1.7 & 4.1905 & -4.0176 & -1.0618 & -9.7700 & 4.3989 & -0.4205 & 0.9899 \\
      1.8 & 4.3249 & -3.9462 & -1.2676 & -9.6627 & 4.1572 & -0.5211 & 1.0905 \\
      1.9 & 4.5209 & -3.8447 & -1.5651 & -9.5140 & 3.8125 & -0.6631 & 1.2325
    \end{tabular}
  \end{ruledtabular}
  \end{minipage}
\end{table*}
The perturbative matching from the lattice to \MSbar\ schemes is
performed by comparing one-loop calculations of quark two-point one
particle irreducible ($1$PI) functions with an insertion of the
relevant bare lattice operator. This requires the evaluation of the
diagrams shown in Fig.~\ref{fig:ptdiags}, together with wavefunction
renormalisation factors, Fig.~\ref{fig:wavefnren}. For the
first-moment operator, we define
\begin{equation}
  \label{eq:Z1ddef}
  \firstmomentop^\MSbar(\mu) = Z_{\firstmomentop}(\mu a)
    \firstmomentop^\latt(a)\,.  
\end{equation}
For the second moment calculation we must take account of mixing with
a total derivative operator (c.f.\ Sec.~\ref{subsec:opmix}). Adopting
the notation
\begin{equation}
  \ODD = \overline{\psi} \gamma_{\{\mu}\gamma_5 \ovlra{D}_\nu
  \ovlra{D}_{\kappa\}} \psi,\quad
  \Odd = \partial_{\{\nu}\partial_\kappa \overline{\psi} \gamma_{\mu\}}
  \gamma_5 \psi,
\end{equation}
with all Lorentz indices distinct and symmetrised, we need to determine
\begin{equation}
  \label{eq:Z2ddef}
  \ODD^\MSbar(\mu) =
  Z_{DD,DD}(\mu a) \ODD^\mathrm{latt}(a) +
  Z_{DD,\partial\partial}(\mu a) \Odd^\mathrm{latt}(a).
\end{equation}
The renormalisation factors are given by
\begin{widetext}
\begin{align}
  \label{eq:Zpt}
  Z_{\firstmomentop}(\mu a) &= \frac1{(1-w_0^2)Z_w}
  \left[ 1 + \bigG \left( -\frac{16}3 \ln(\mu a) + \Sigma_1^\MSbar
      - \Sigma_1 + V^\MSbar - V \right) \right]\,,\\
  \label{eq:ZDDDD}
  Z_{DD,DD}(\mu a) &= \frac1{(1-w_0^2)Z_w}
  \left[ 1 + \bigG \left( -\frac{25}3 \ln(\mu a) + \Sigma_1^\MSbar
      - \Sigma_1 + \VDD^\MSbar - \VDD \right) \right], \\
  \label{eq:ZDDdd}
  Z_{DD,\partial\partial}(\mu a) &= \frac1{(1-w_0^2)Z_w} \,
  \bigG \left( \frac53 \ln(\mu a) + \Vdd^\MSbar - \Vdd \right).
\end{align}
\end{widetext}
In the equations above $(1-w_0^2)Z_w$ is a characteristic
normalisation factor for the physical quark fields in the domain-wall
formalism. $Z_w$ represents an additive renormalisation of the large
Dirac mass or domain-wall height $M=1-w_0$, which can be rewritten in
multiplicative form at one-loop as
\begin{equation}
  Z_w = 1+\bigG \,z_w,
  \qquad
  z_w = \frac{2w_0}{1-w_0^2}\,\Sigma_w.
\end{equation}
The one-loop correction $z_w$ becomes very large for certain choices of
$M$~\cite{Aoki:1998vv, Aoki:2002iq}, including that used in our numerical
simulations, so that some form of mean-field improvement is necessary, as
discussed below.

Terms with superscripts \MSbar\ in Eqs.~\eqref{eq:Zpt},
\eqref{eq:ZDDDD} and~\eqref{eq:ZDDdd} arise from the continuum
calculations, while unsuperscripted terms come from the
computations in the lattice scheme. To shorten some expressions below
we will define
\begin{align}
  \label{eq:c}
  c &= \Sigma_1^\MSbar - \Sigma_1 + V^\MSbar - V, \\
  \label{eq:cDD}
  \cDD &= \Sigma_1^\MSbar - \Sigma_1 + \VDD^\MSbar - \VDD, \\
  \label{eq:cdd}
  \cdd &= \Vdd^\MSbar - \Vdd.
\end{align}
The terms $\Sigma_1^\MSbar$ and $\Sigma_1$ come from quark wavefunction
renormalisation, while $V^\MSbar$, $\VDD^\MSbar$, $\Vdd^\MSbar$ and $V$, $\VDD$,
$\Vdd$ come from the one-loop corrections to the amputated two-point function.
They are given by ``vertex'' and ``sail'' diagrams, plus an operator tadpole
diagram in the lattice case. $\VDD^\MSbar$ and $\VDD$ can be isolated by
computing the one-loop correction with equal incoming and outgoing quark
momenta. Likewise $\Vdd^\MSbar$ and $\Vdd$ are found by setting the incoming and
outgoing quark momenta equal and opposite (the lattice tadpole diagram does not
contribute in this case). Using naive dimensional regularisation (NDR) in
Feynman gauge with a gluon mass infrared (IR) regulator,
\begin{align}
  \Sigma_1^\MSbar &= \frac12,
  &
  V^\MSbar &= -\frac{25}{18},\\
  \VDD^\MSbar &= -\frac{121}{72},
  &
  \Vdd^\MSbar &= \frac{41}{72}.
\end{align}
The lattice contributions are evaluated for domain-wall fermions with
the Iwasaki gluon action ($c_1=-0.331$), also choosing Feynman gauge
and using a gluon mass IR regulator. $\Sigma_1$ has been evaluated
in~\cite{Aoki:2002iq}, while we calculated the vertex term $V$ for the
first moment operator in~\cite{Boyle:2006pw}. Here we have calculated
the vertex terms $\VDD$ and $\Vdd$ for the second moment operator.
Perturbative calculations with domain-wall fermions are explained
in~\cite{Aoki:1998vv, Aoki:2002iq} and the form of the Iwasaki gluon
propagator can be found in~\cite{Iwasaki:1983ck}. Values for
$\Sigma_1$, $V$, $\VDD$ and $\Vdd$ are given as functions of $M$ in
Table~\ref{tab:VvsM}, along with $c$, $\cDD$ and $\cdd$. Chiral
symmetry of the domain-wall action implies that these results also
apply for the operators which are like those used here, but without
the $\gamma_5$. We note that the perturbative renormalisation factor
for the first moment operator using alternative fermion and gauge
formulations can be found in~\cite{Capitani:2005vb} (domain-wall
fermions and plaquette action),~\cite{Horsley:2005jk} (overlap
fermions and L\"uscher--Weisz action) and~\cite{Gockeler:2006nb}
(clover fermions and plaquette action). Second moment calculations
with clover and Wilson fermions have been performed
in~\cite{Gockeler:2006nb} and~\cite{Gockeler:2004xb} respectively (in
both cases using the plaquette action).

\begin{table}
  \caption{Values for $z_w$, $z_w^\mathrm{MF}$ extracted from the
    results in~\cite{Aoki:2002iq}, and $d_f$ extracted
    from~\cite{Aoki:2003uf}.}
  \label{tab:zwmf}
  \begin{ruledtabular}
    \begin{tabular}{D..1D..4D..4D..7}
      \multicolumn1c{$M$} &
      \multicolumn1c{$z_w$} &
      \multicolumn1c{$z_w^\mathrm{MF}$} &
      \multicolumn1c{$d_f$}\\
      \hline
      0.1 & -243.86 & -86.579 & -0.02303\\
      0.2 & -113.29 & -39.501 & -0.01798\\
      0.3 & -69.404 & -23.830 & -0.01497\\
      0.4 & -47.077 & -15.949 & -0.01274\\
      0.5 & -33.278 & -11.142 & -0.01090\\
      0.6 & -23.648 & -7.8365 & -0.009315\\
      0.7 & -16.300 & -5.3538 & -0.007896\\
      0.8 & -10.263 & -3.3459 & -0.006589\\
      0.9 & -4.9617 & -1.6078 & -0.005379\\
      1.0 &  0.0   &  0.0   & -0.004261\\
      1.1 & 4.9442 & 1.5902 & -0.003227\\
      1.2 & 10.192 & 3.2748 & -0.002290\\
      1.3 & 16.136 & 5.1900 & -0.001485\\
      1.4 & 23.346 & 7.5350 & -0.0008650\\
      1.5 & 32.784 & 10.648 & -0.0005360\\
      1.6 & 46.322 & 15.194 & -0.0006566\\
      1.7 & 68.294 & 22.720 & -0.001570\\
      1.8 & 111.69 & 37.901 & -0.004014\\
      1.9 & 241.55 & 84.270 & -0.01020
    \end{tabular}		
  \end{ruledtabular}
\end{table}
Our numerical simulations use $M=1.8$. For this value of $M$, with the
Iwasaki gluon action, the one-loop coefficient in the physical quark
normalisation is $z_w \approx 112$ (extracted from $\Sigma_w$ in
Table~III of~\cite{Aoki:2002iq}), making it clear that mean-field
improvement is necessary. We follow the prescription given
in~\cite{Aoki:2002iq}. The first step is to define a mean-field value
for the domain-wall height,
\begin{equation}
  \Mmf = M - 4(1-P^{1/4}) = 1.3029
\end{equation}
where $P=0.58813(4)$ is the average plaquette value in the chiral
limit in our simulations. The physical quark normalisation factor
becomes $\left[1-(\wmf)^2\right]Z_w^\mathrm{MF}$, with
\begin{equation}
\begin{aligned}
  \label{eq:Zzwmf}
  Z_w^\mathrm{MF} &= 1+\bigG z_w^\mathrm{MF},\\
  z_w^\mathrm{MF} &= \frac{2\wmf}{1-(\wmf)^2}\,(\Sigma_w + 32\pi^2\Tmf)
  = 5.2509,
\end{aligned}
\end{equation}
where $\Tmf=0.0525664$~\cite{Aoki:2002iq} is a mean-field tadpole
factor and $\Sigma_w$ is evaluated at $\Mmf$. Values for
$z_w^\mathrm{MF}$ as a function of $M$ are quoted in
Table~\ref{tab:zwmf}, extracted from the results
in~\cite{Aoki:2002iq}. Likewise, $\Sigma_1=3.9731$, $V=-4.1907$,
$\VDD=-10.045$ and $\Vdd=-0.1696$ are evaluated at $\Mmf$.

For the operator $\ODD$ with two covariant derivatives, mean-field improvement
introduces a factor $u_\mathrm{pt}/u$ where $u$ is the mean link (here taken to
be $u=P^{1/4}$) and
\begin{equation*}
  u_\mathrm{pt}=1-\bigG\,8\pi^2\Tmf
\end{equation*}
is its perturbative expansion. For $\Odd$ with two ordinary
derivatives, in contrast, the extra factor is $u/u_\mathrm{pt}$. The
mean-field-improved matching factors are thus
\begin{widetext}
\begin{align}
  \label{eq:Z-MF}
  Z_{\firstmomentop}^\mathrm{MF}
  &= \frac1{1-(\wmf)^2}\,\frac1{Z_w^\mathrm{MF}}
  \left[1+\bigG \left( -\frac{16}3 \ln(\mu a) + c^\mathrm{MF}
    \right) \right]\\
  \label{eq:ZDDDD-MF}
  Z_{DD,DD}^\mathrm{MF}
  &= \frac1u\,\frac1{1-(\wmf)^2}\,\frac1{Z_w^\mathrm{MF}}
  \left[ 1 + \bigG \left( -\frac{25}3 \ln(\mu a) +
      \cDD^\mathrm{MF} - 8\pi^2\Tmf
    \right) \right]\\
  \label{eq:ZDDdd-MF}
  Z_{DD,\partial\partial}^\mathrm{MF}
  &= u\,\frac1{1-(\wmf)^2}\,\frac1{Z_w^\mathrm{MF}} \,
  \bigG \left( \frac53 \ln(\mu a) + \cdd^\mathrm{MF}
  \right)
\end{align}
\end{widetext}
with $c^\mathrm{MF} = -0.6713$, $\cDD^\mathrm{MF} - 8\pi^2\Tmf =
0.7408$ and $\cdd^\mathrm{MF} = 0.7391$. To evaluate these
expressions, we make two choices for the coupling. The first is a
mean-field improved coupling defined using the measured plaquette
value $P$, according to~\cite{Aoki:2003uf}
\begin{multline}
  \label{eq:meas-plaq}
  \frac1{g^2_\mathrm{MF}(\mu)} =
  \frac P{g_0^2} + d_g + c_p +\frac{22}{16\pi^2}\,\ln(\mu a)\\
  + N_f \left[ d_f -\frac4{48\pi^2}\,\ln(\mu a)\right]
\end{multline}
where $N_f$ is the number of dynamical quark flavours. For the Iwasaki
gauge action with $c_1=-0.331$, the values $d_g=0.1053$ and
$c_p=0.1401$ are given in~\cite{Aoki:2002iq}, while values for $d_f$
as a function of $M$ were calculated in~\cite{Aoki:2003uf} and are
quoted in Table~\ref{tab:zwmf}. In our simulations, $\beta = 6/g_0^2 =
2.13$ with $N_f=3$ and $a^{-1}=1.729\gev$. The second choice is the
continuum \MSbar\ coupling, calculated as outlined in Appendix A
of~\cite{Aoki:2007xm}. At $\mu a=1$, we find $\alpha_\mathrm{MF} =
0.1769$ and $\alpha^\MSbar = 0.3138$. We use these two values to
evaluate the renormalisation factors above. We also evaluate the
mean-field improved expression for the axial vector current
renormalisation~\cite{Aoki:2002iq}, interpolating to our mean-field
$\Mmf$. The values are shown in Table~\ref{tab:ptrenorm}. The ratios
of the renormalisation factors, from which the factor
$1/(1-(\wmf)^2)Z_w^\mathrm{MF}$ cancels, are also shown in the table.
\begin{table*}
\caption{Perturbative renormalisation factors and their ratios for two
  choices of the strong coupling, evaluated at $\mu a = 1$.}
\label{tab:ptrenorm}
\begin{ruledtabular}
\begin{tabular}{lccccccc}
  & $Z_{\firstmomentop}^\mathrm{MF}$ & $Z_{DD,DD}^\mathrm{MF}$
    & $Z_{DD,\partial\partial}^\mathrm{MF}$ & $Z_\mathrm{A}^\mathrm{MF}$
    & $\frac{Z_{\firstmomentop}^\mathrm{MF}}{Z_\mathrm{A}^\mathrm{MF}}$
    & $\frac{Z_{DD,DD}^\mathrm{MF}}{Z_\mathrm{A}^\mathrm{MF}}$
    &
  $\frac{Z_{DD,\partial\partial}^\mathrm{MF}}{Z_\mathrm{A}^\mathrm{MF}}$\\[1.1ex]
\hline
$\alpha_\mathrm{MF}$ & 0.9896 & 1.1604 & 0.0122 & 0.8009 &
                    1.2356 & 1.4488 & 0.0152\\
$\alpha^\MSbar$       & 0.9162 & 1.0966 & 0.0202 & 0.6934 & 
                     1.3214 & 1.5815 & 0.0291
\end{tabular}
\end{ruledtabular}
\end{table*}

We take the mean value of the results with the two different choices
for the coupling as the best answer for the renormalisation factors.
The difference between the two choices will form the error. The
relevant factors for the perturbative renormalisation of the ratios in
Eqs.~\eqref{eq:Pratio} and~\eqref{eq:Vratio} are given in
Table~\ref{tab:PTZs}. Chiral symmetry here ensures that we do not have
to distinguish between vector and axial-vector operators. We note that
the contribution from the mixing term $Z_{DD,\partial\partial}$ is
smaller than the error on $Z_{DD,DD}$ itself.
\begin{table}
  \caption{Perturbative renormalisation factors to match the
    lattice results to \MSbar\ at $a\mu = 1$.}
  \label{tab:PTZs}
  \begin{minipage}{.5\hsize}
  \begin{ruledtabular}
    \begin{tabular}{D..5D..5D..6}
      \multicolumn1c{$\frac{Z_{\firstmomentop}}{Z_\mathrm{A}}$} &
      \multicolumn1c{$\frac{Z_{DD,DD}}{Z_\mathrm{A}}$} &
      \multicolumn1c{$\frac{Z_{DD,\partial\partial}}{Z_\mathrm{A}}$}\\[1.1ex]
      \hline
      1.28(4) & 1.52(7) & 0.022(7)\\
    \end{tabular}		
  \end{ruledtabular}
  \end{minipage}
\end{table}

\subsection{Nonperturbative Renormalisation}
\label{sec:NPR}

In order to renormalise the correlation functions nonperturbatively
we make use of the Rome-Southampton
\ripm\ scheme~\cite{Martinelli:1994ty} which we now briefly review and
discuss refinements to~\cite{Aoki:2007xm}. The starting point and
definition of the \ripm\ scheme is a simple renormalisation condition
that can be imposed independently of the regularisation used, thus on
the lattice as well as in the continuum. This facilitates scheme
changes which is important for the matching to \MSbar. The
renormalisation condition has the form
\begin{equation}
  \label{eq:npr:rencond}
   \Lambda_\op{O}(p) = Z_\op{O}(\mu) Z_q^{-1}(\mu)\,
     \Lambda^\text{bare}_\op{O}(p)\Bigr|_{p^2=\mu^2}
   = \Lambda^\text{tree}_\op{O}(p),
\end{equation}
where $\Lambda_\op{O}$ ($\Lambda^\text{bare}_\op{O}$) is the
renormalised (bare) vertex amplitude. Together with the quark field
renormalisation $Z_q^{1/2}$, defined by $\psi =
Z_q^{1/2}\psi^\mathrm{bare}$, this defines the renormalisation
constant $Z_\op{O}$ for the operator $\op{O}$. The renormalisation
scale $\mu$ is set by the momentum of the external states entering the
vertex amplitude. In the original \ripm\ scheme these momenta are
\emph{exceptional}, that is equal incoming and outgoing quark momenta,
$p$ and $p'$. For some renormalisation factors it is advantageous to
use a non-exceptional symmetric choice of momenta $p^2=p'^2=q^2$,
where $q=p-p'$, leading to the distinct \rism\ scheme. This suppresses
unwanted infrared effects in the vertex amplitude, pion poles for
example, and suggests a better-behaved accompanying continuum
perturbation theory~\cite{Sturm:2009kb}. Exceptional momenta with
$q=0$ also cause matrix elements of operators with total derivatives
to vanish, making $Z_{DD,\partial\partial}$ inaccessible in our
nonperturbative analysis.

The vertex amplitude is constructed from the unamputated Green's
function
\begin{equation}
  \label{eq:npr:unamp}
\begin{split}
  G_\op{O}(p) &= \bret{\psi(p) \op{O}(0) \overline{\psi}(p)}\,,\\
  \op{O}(0) &= \sum_{x,x'} \bar{\psi}(x) J_\op{O}(x,x') \psi(x')\,.
\end{split}
\end{equation}
The external quark lines need gauge fixing, for which we use Landau
gauge. The current $J$ has the appropriate Dirac structure and may be
non-local if the operator contains derivatives. For example, a single
right derivative $\ovra{D}_\nu$ in the vector case would correspond to
\begin{equation}
  \label{eq:npr:J}
  J_{\op{O}_{\rho\mu}}(x,x') = \gamma_\rho \, \frac{1}{2} \Bigl(
  U(x,x') \delta_{x',x+\hat{\mu}} - U(x,x') \delta_{x',x-\hat{\mu}}
  \Bigr)
\end{equation}
matching the definition in Eq.~\eqref{eq:derdef}.

The vertex amplitude itself is found after amputating the Green's
function and tracing with a suitable projector $P_\op{O}$
\begin{align}
  \label{eq:npr:amproj}
  \Lambda_\op{O}(p) &= \Tr \left[ \Pi_\op{O}(p) P_\op{O} \right]\\
  \intertext{with}
  \Pi_\op{O}(p) &= \bret{S(p)}^{-1} \bret{G_\op{O}(p)} \bret{S(p)}^{-1}\,.
\end{align}
We have used the quark propagator $S(p)$ and the angle brackets
indicate the gauge average. The projector $P_\op{O}$ depends on the
particular operator and includes an overall normalisation factor to
account for the colour and Dirac trace. In a simple example $P_\op{O}$
would isolate the tree-level contribution to the vertex amplitude; we
will detail our choices below. We have now defined the renormalisation
procedure and will turn to details of the implementation before
discussing the results.

\subsubsection{Momentum sources}

One refinement to our previous work~\cite{Aoki:2007xm} is the use of
momentum sources~\cite{Gockeler:1998ye}. In contrast to the point
sources used before, this effectively amounts to a volume average over
the lattice resulting in much smaller statistical
errors~\cite{Boyle:2008nj}. Starting from \eqref{eq:npr:unamp} the
Green's function in momentum space is
\begin{equation}
  \label{eq:npr:pspace}
  G_\op{O}(p) =
  \sum_{x,x'} \bret{
    \gamma_5 S^\dagger(p)_x \gamma_5\,
    J_\op{O}(x,x')\,S(p)_{x'} }\,,
\end{equation}
where rather than use the quark propagator $S(x|y)$ obtained by
inverting the Dirac Matrix $M$ on a point source
\begin{equation}
  \label{eq:npr:prop}
  \sum_{x} M(x',x)S(x|y) = \delta_{x',y}\,.
\end{equation}
we use
$S(p)_x = \sum_y S(x|y) e^{\I py}$ which can be found by inverting
with a momentum source~\cite{Gockeler:1998ye}
\begin{equation}
  \label{eq:npr:pprop}
  \sum_x M(x',x)S(p)_{x} = e^{\I px'}\,,
\end{equation}
and is defined on all lattice sites corresponding to the off-shell
quarks used in the Green's function. The gain in statistical accuracy
is paid for with a separate inversion for every momentum used in the
simulation. However, this is more than compensated by a much reduced
number of necessary configurations. Limiting ourselves to a few
carefully chosen momenta, statistical fluctuations are reduced with
lower overall computational cost.

The momenta we use are first of all constrained to be within a range
$\Lambda_\text{QCD}\ll p^2 \ll 1/a$ for the
\ripm\ scheme~\cite{Martinelli:1994ty}. We use our previous
results~\cite{Aoki:2007xm} to identify suitable values and focus on
momenta which are expected to have reduced hypercubic lattice
artefacts by trying to limit $\sum p_\mu^4$ for fixed
$p^2$~\cite{Boyle:2008nj} (see
also~\cite{Boucaud:2003dx,deSoto:2007ht}). The values used are:
\begin{align*}
  16^3 \times 32:
   &\quad(1,1,2,3),\,(1,1,2,4),\,(1,2,2,1),\\
   &\quad(1,2,2,3),\,(1,2,2,4)\\
\intertext{and}
  24^3 \times 64:
   &\quad(2,2,2,7),\,(2,2,2,8),\,(2,2,3,7),\\
   &\quad(2,2,3,8),\,(2,3,3,7) ,
\end{align*}
where we have given $n_\mu^\tra$ for momenta $p_\mu=2\pi n_\mu/L$
(with $L\to T$ for time components).

\subsubsection{Projectors}

We extend the set of operators considered previously
in~\cite{Aoki:2007xm}. We now require operators with up to two
derivatives, $\op{O}^{(5)}_{\{\mu_1\dots\mu_n\}}$ $(n\leq3)$, making
the the necessary projectors $P_\op{O}$ slightly more involved than
for bilinears. Since we resort to readily available
calculations~\cite{Gracey:2003yr, Gracey:2003mr, Gracey:2006zr} for
the final conversion to \MSbar\ as well as to account for running, we
have to tailor the projectors to match the \ripm\ scheme and vertex
functions used in the continuum calculations. Decomposing the
amputated Green's function into terms allowed by Lorentz symmetry and
remembering that we are taking all indices to be distinct, we
find~\cite{Gracey:2003yr,Gracey:2003mr,Gracey:2006zr}
\begin{equation}
  \label{eq:npr:grdeco}
  G^{}_\op{O}(p) = 
  \Sigma_1(p) \,\gamma_{\{\mu_1} p_{\mu_2} \dots p_{\mu_n\}} +
  \Sigma_2(p) \,p_{\mu_1} \dots p_{\mu_n} \slashed{p}\,.
\end{equation}
For simplicity we limit the discussion to the vector case here;
axial-vector operators are analogous. The \ripm\ scheme uses the
contribution from $\Sigma_1(p)$ only in \eqref{eq:npr:grdeco}. The
required projector $P_\op{O}$ will depend on the momentum entering the
Green's function and its (fixed) directions $\mu_i$ $(i=1\dots n)$. In
general, multiplying $G_\op{O}$ with $\gamma_{\mu_i}$ picks up
combinations of both terms $\Sigma_1$ and $\Sigma_2$. On the other
hand, projecting with $\gamma_\rho$ where $\rho \notin \{\mu_i\}$ is
only sensitive to $\Sigma_2$ (note that we have $n \leq 3$). Thus
multiplying with the difference of the two Dirac matrices with
appropriate normalisation and momentum factors ensures that the vertex
amplitude in \eqref{eq:npr:amproj} contains $\Sigma_1(p)$ only. There
are simpler special cases in which one or more components of the
momentum $p$ are zero, causing the second term in
\eqref{eq:npr:grdeco} to vanish. However, since we tried to choose our
momentum directions close to the diagonal of the lattice, we do not
have momentum components that are zero.

For fixed indices $\mu_i$ $(i=1\dots n)$ of the Green's function we
can construct $n$ different projectors $P_{\op{O},i}$ by starting from
any of the $\gamma_{\mu_i}$:
\begin{equation}
  \label{eq:npr:projs}
  P_{\op{O},i} =
  \frac{\displaystyle
  \gamma_{\mu_i} - \gamma_\rho\frac{\bar{p}_{\mu_i}}{\bar{p}_\rho}}
  {\displaystyle
   \mathcal{N}\,\prod^n_{j\ne i,j=1} \bar{p}_{\mu_j}}
  \,,  \text{ with }i=1\dots n\,.
\end{equation}
The normalisation $\mathcal{N}$ is chosen such that for the tree-level
vertex amplitude we find $\Lambda_\op{O}^\text{tree}(p) = 1$. The
index $\rho$ is different from any of the $\mu_i$ and such that its
momentum component $\bar{p}_\rho$ is as small as possible to reduce
discretisation errors. We use $\bar{p}_\mu = \sin p_\mu$ to better
account for lattice momenta. The case of axial-vector operators
$\op{O}^5$ is straightforward, with $\gamma_5$ inserted in the
appropriate places.

Combining the $n$ different $P_{\op{O},i}$ with the possible index
combinations of the Green's functions results in a total of $4$, $12$
and $12$ $(n=1,2,3)$ choices to compute the vertex amplitude
$\Lambda_\op{O}(p)$ in Eq.~\eqref{eq:npr:amproj}, all of which should
provide the same result for the final renormalisation constant in the
absence of lattice artefacts. Because of the different sized momentum
components in different lattice directions, the expected
discretisation errors vary depending on the directions selected by the
indices of the projector. We reflect these artefacts coming from
breaking continuum O(4) symmetry to lattice hypercubic symmetry in the
systematic error of our final results. With additional lattice
spacings and the use of partially-twisted boundary conditions, we
could eliminate hypercubic lattice artefacts in the continuum
limit~\cite{Arthur:2010ht,Arthur:2010hy}.

\subsubsection{Quark field renormalisation}

In general, the renormalisation condition Eq.~\eqref{eq:npr:rencond}
requires knowledge of the field renormalisation $Z_q$ to obtain
$Z_\op{O}$. However, in the present calculation only ratios of
renormalisation factors of operators with one, two or no derivatives
appear, Eqs.~\eqref{eq:Pratio} and \eqref{eq:Vratio}. Combining this
with our renormalisation condition leads to,
\begin{equation}
  \label{eq:npr:zratio}
  \frac{Z_{\op{O},n=2,3}(\mu)}{Z_{\op{O},n=1}(\mu)} =
  \left.\frac{\Lambda^\text{bare}_{\op{O},n=1}(p)}
             {\Lambda^\text{bare}_{\op{O},n=2,3}(p)}\right|_{p^2=\mu^2}\,,
\end{equation}
where the explicit $Z_q$ dependence drops out. As mentioned earlier,
we can use either the vector or axial-vector bilinears in this ratio
thanks to chiral symmetry. We follow our earlier
procedure~\cite{Aoki:2007xm} and average $\Lambda_{\gamma_\rho}$ and
$\Lambda^{(5)}_{\gamma_\rho}$ ($\Lambda_V$/$\Lambda_A$ in the
reference) to obtain our best answer. The analysis is also performed
with $\Lambda^{(5)}_{\gamma_\rho}$ only and the difference of the two
enters our systematic error.

\subsubsection{Results for renormalisation factors}

\begin{figure}
  \includegraphics[width=\hsize,bb=89 513 298 665]{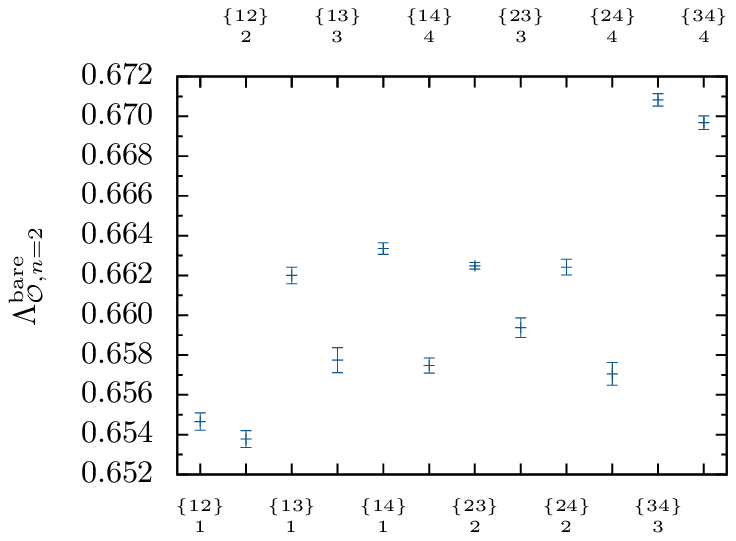}\\[1ex]
  \includegraphics[width=\hsize,bb=89 513 298 665]{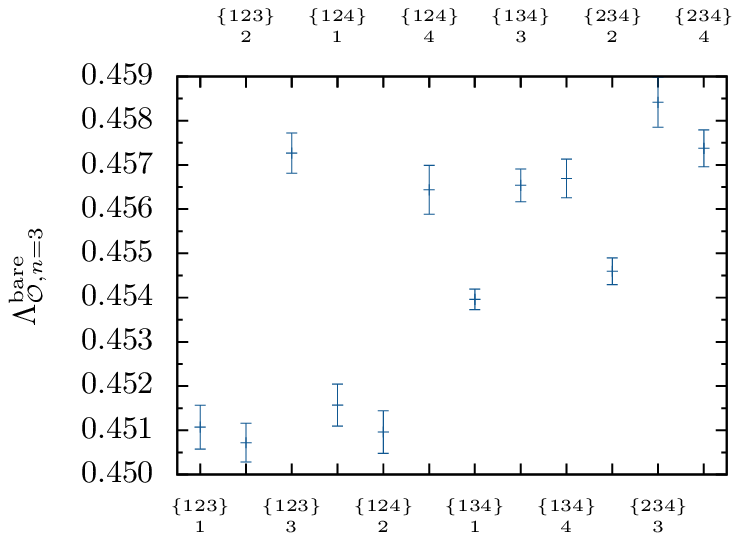}
  \caption{Results for $\Lambda^\text{bare}_{\op{O},n=2}$
    ($\Lambda^\text{bare}_{\op{O},n=3}$) on the top (bottom) for a
    fixed momentum $(ap)^2 = 1.78201$, $p^\tra=(2,2,3,8)$. The labels
    above and below the plots show the indices of the Green's function
    $\{\mu_i\}$ (top) and projector (bottom). The disagreement between
    the different projections is due to lattice artefacts.}
  \label{fig:npr:disc}
\end{figure}

Compared to~\cite{Aoki:2007xm} the reduced statistical errors make
previously hidden systematic effects apparent and
quantifiable~\cite{Boyle:2008nj} and affect the way we extract the
renormalisation factors. We start by considering different projectors
for a fixed momentum $p_\mu$ of the external quarks, see
Fig.~\ref{fig:npr:disc}. The results should be independent of the
rotation and size of the momentum components used for the projector.
The smaller statistical errors now reveal a disagreement due to
lattice artefacts. We combine all choices for our best answer and
account for the spread in our systematic error, improving previous
estimates.
Our general recipe to obtain the renormalisation factors follows. The
ratio of bare vertex amplitudes is extrapolated linearly to the chiral
limit $m_q \to -m_\text{res}$ for each momentum. Only in the chiral
limit can we remove the running of our data points and match them to a
continuum scheme. So by using~\cite{Gracey:2003yr, Gracey:2003mr,
  Gracey:2006zr} we take our results from the \ripm\ scheme at scale
$\mu^2=p^2$ to a common scale $\mu^2=4\gev^2$ and convert to
\MSbar\ at that scale. The values thus obtained are then linearly
interpolated to $p^2 = (2\gev)^2$ within our momentum window to obtain
$Z_{\op{O},n=2,3}/Z_{\op{O},n=1}$ at a scale $\mu=2\gev$.

The central value is computed from the averaged values from all
projectors and index combinations. A standard bootstrap analysis
provides the statistical error which is inflated with
$\sqrt{\chi^2/\text{d.o.f.}}$ (the PDG scale-factor~\cite{pdg2010})
from the interpolation. Several effects are taken into account for the
systematic error. Lattice artefacts are the dominant effect. To
estimate those, we perform the analysis for all projectors separately
as indicated above and chose the highest and lowest result for each
momentum for the interpolation. From the two fits, the larger
deviation from the central value then constitutes the systematic error
from discretisation effects (labelled `spread' in the final table).
This is a conservative approach for the discretisation error. Taking
random choices of projectors (or rather their direction) for each
momentum and looking at the $1\sigma$ width of the range of results
for many of those picks would lead to a smaller error. We account for
missing higher order terms in the continuum perturbative calculation
via the slope of the momentum interpolation, using the difference of
our results at $p^2 = (2\gev)^2$ and $(0\gev)^2$, indicated by
`slope', as a measure. We note, however, that we cannot disentangle
perturbative and discretisation errors here and thus double count some
of the discretisation effects. Another source of systematic error is
the strange quark mass, kept fixed at $\ms=0.04$ in our simulation. We
deal with that as described at the end of section IV.F
in~\cite{Aoki:2007xm}, estimating an error from half the linear
dependence (slope) multiplied by the strange quark mass, $m_\text{s}$.
This error is labelled `$\Delta m_s$'. The last contribution to the
systematic error is from the chiral symmetry breaking evident when
comparing our vector and axial-vector operators~\cite{Aoki:2007xm,
  Aoki:2009ka, Sturm:2009kb}. This is estimated by the difference of
the final results when taking the axial-vector bilinear $(n=1)$ or the
averaged vector and axial-vector bilinear for the ratio in
Eq.~\eqref{eq:npr:zratio} (labelled `$V-A$'). Adding the four
contributions in quadrature gives our systematic error.

\begin{figure}
  \includegraphics[width=0.95\hsize,
                   bb=89 513 303 657]{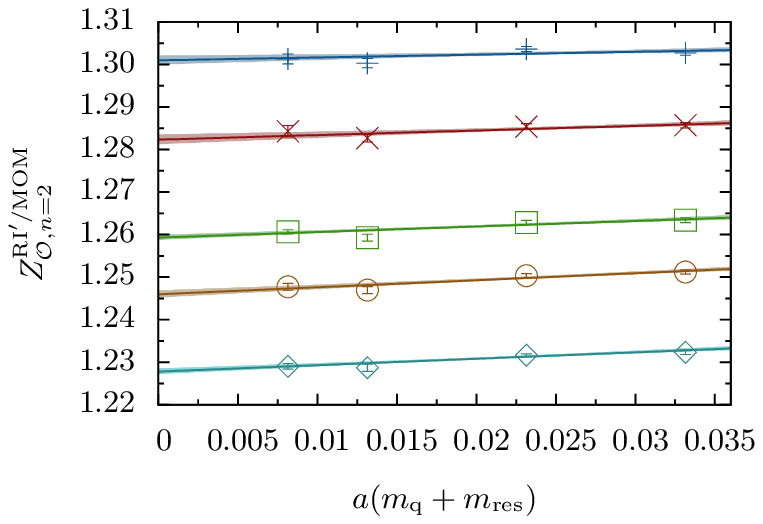}\\[1ex]
  \includegraphics[width=0.95\hsize,
                   bb=89 513 303 657]{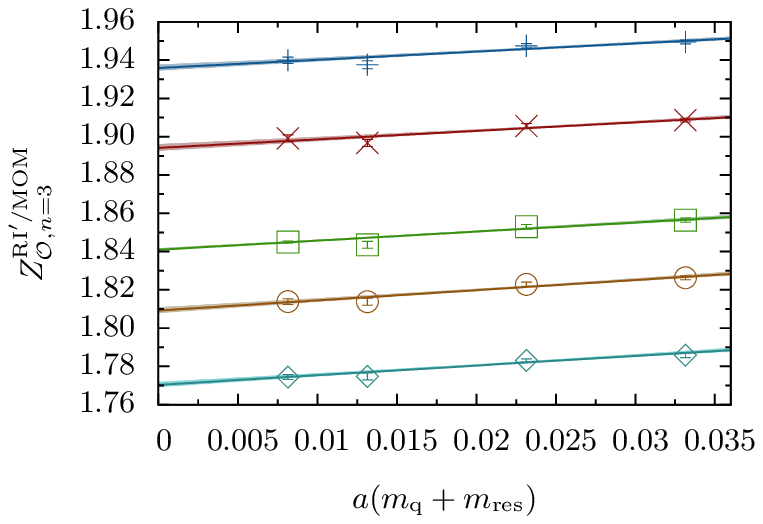}
  \caption{Linear extrapolations of the renormalisation factors to the chiral
    limit. The top shows $Z^\ripm_{\op{O},n=2}$, the bottom plot is for
    $Z^\ripm_{\op{O},n=3}$. The momenta are increasing from top to bottom and we
    have $(ap)^2=1.2947,1.4392,1.6374,1.7820$ and $1.9801$.}
  \label{fig:npr:chirlim}
\end{figure}

\begin{figure}
  \includegraphics[width=0.93\hsize,
                   bb=89 511 298 654]{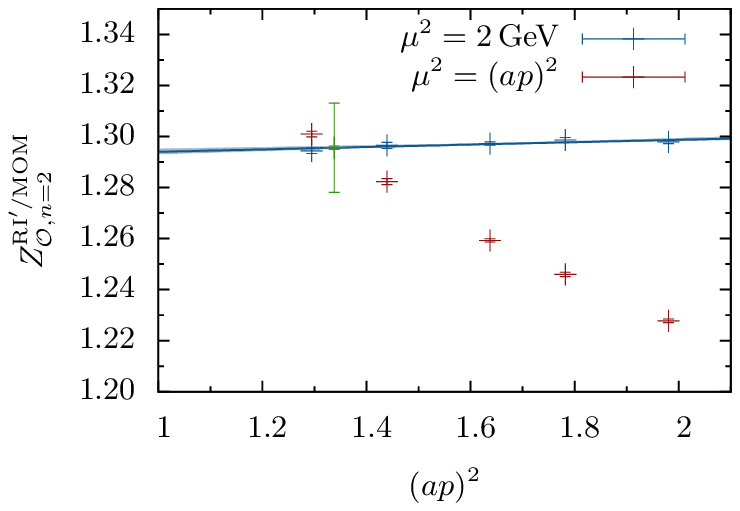}\\[1ex]
  \includegraphics[width=0.93\hsize,
                   bb=89 511 298 657]{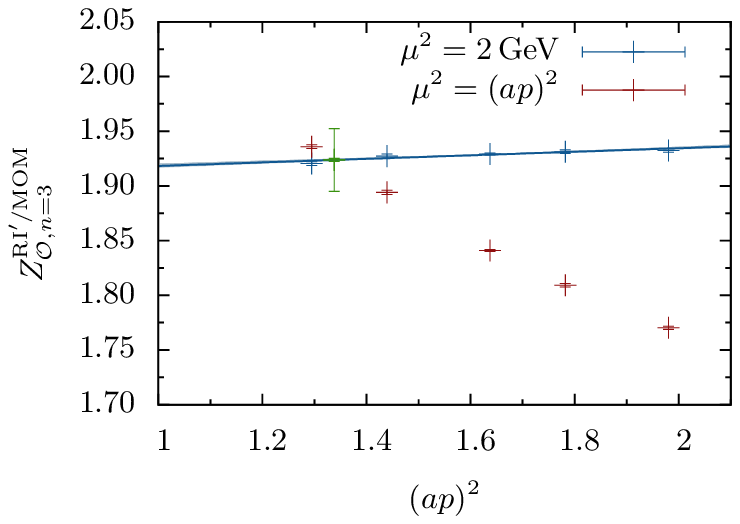}
  \caption{These plots show the scale dependent $Z$'s and $Z$'s for a
    fixed scale $\mu=2\gev$ with the running successfully removed
    (both in \ripm). The top (bottom) plot is for the one (two)
    derivative case. Also included are the linear interpolation to the
    final result with the statistical error indicated by the error
    band. Our result at $\mu=2\gev$ is then shown with error bars for
    the statistical and systematic errors.}
  \label{fig:npr:scale}
\end{figure}

To illustrate some of the steps mentioned above, we include in
Fig.~\ref{fig:npr:chirlim} two examples of the extrapolation to the
chiral limit. Shown are extrapolations for all our five momenta, for
the renormalisation factors for one and two derivatives.
In Fig.~\ref{fig:npr:scale} we show the renormalisation factors before
and after we remove the running, again for one and two derivatives.
Once at a common scale, the data points are much flatter indicating
the validity of the scale conversion and the momentum window. Also
included is the linear fit and our final result.

Our final values for the ratios of renormalisation factors are given
in Table~\ref{tab:npr:res}. These have been obtained using vector like
operators. Results from the axial-vector operators are almost
identical and show no low energy effects from breaking chiral
symmetry, as for bilinears. We note that the renormalisation factors
are significantly different from one and deviate substantially from
the perturbative results. Thus nonperturbative renormalisation looks
imperative here.
\begin{table*}
 \caption{Final results for the renormalisation factors in \MSbar\ at
   $\mu=2\gev$. Results are given for both lattice sizes with all
   systematic errors. The perturbative results are also shown for
   comparison.}
 \label{tab:npr:res}
 \begin{minipage}{.7\textwidth}
 \begin{ruledtabular}
 \begin{tabular}{l|cc|cc}
  & \multicolumn{2}{c|}{$Z_{\op{O}_{\{\rho\nu\}}}/Z_A$} & 
    \multicolumn{2}{c}{$Z_{DD,DD}/Z_A$} \\
  & $16^3\times 32$ & $24^3\times 64$ & $16^3\times 32$ & $24^3\times 64$  \\ 
\hline
central value       & 1.54575 &   1.52893    & 2.06064 &   2.02800     \\
statistical error   & 0.00249 &   0.00081    & 0.00482 &   0.00149     \\
spread              & 0.02968 &   0.01809    & 0.03702 &   0.01534     \\
slope               & 0.00470 &   0.00743    & 0.00097 &   0.02285     \\
$\Delta m_s$        & 0.00089 &   0.00232    & 0.00469 &   0.00992     \\
$V-A$               & 0.00723 &   0.00602    & 0.00938 &   0.00760     \\
total error         & 0.03102 &   0.02061    & 0.03879 &   0.03026     \\
\hline
\multicolumn{2}{l}{best result} & 1.5289(8)(206)   & & 2.028(1)(30)     \\
\hline
\multicolumn{2}{l}{perturbative result} & 1.24(3)  & & 1.45(5)
 \end{tabular}
 \end{ruledtabular}
 \end{minipage}
\end{table*}

\subsection{Renormalised Results}
\label{sec:renresults}

We now use the renormalisation factors from the previous section to
convert our bare lattice results to \MSbar\ at $\mu=2\gev$. The local
matrix elements in Eq.~\eqref{eq:dadef} require the renormalisation
factors $Z_{\firstmomentop}$, $Z_{DD,DD}$ and
$Z_{DD,\partial\partial}$ as defined in Eqs.~\eqref{eq:Z1ddef} and
\eqref{eq:Z2ddef}. The first two are computed nonperturbatively while
for the last one we use the perturbative result. From
Eq.~\eqref{eq:Z2ddef} we see that the mixing term requires the
computation of a matrix element with an operator insertion of $\Odd$.
This is simplified since we use the ratios \eqref{eq:Pratio} and
\eqref{eq:Vratio} to extract the moments of the distribution
amplitudes. Within the ratios, the matrix element with the operator
$\Odd$ differs from the denominator only by the momentum factors and
thus does not have to be computed separately. It contributes a
constant shift to the result. To summarise:
\begin{subequations}
  \label{eq:resum}
  \begin{align}
    \label{eq:resum1}
    \bret{\xi^1}^\MSbar &= \frac{Z_{\firstmomentop}}{Z_A}
    \bret{\xi^1}^\text{bare}\,,\\
    \label{eq:resum2}
    \bret{\xi^2}^\MSbar &= \frac{Z_{DD,DD}}{Z_A} \bret{\xi^2}^\text{bare} +
    \frac{Z_{DD,\partial\partial}}{Z_A}\,.
  \end{align}
\end{subequations}
With our best nonperturbative results from Table~\ref{tab:npr:res} and the
perturbative result for the mixing term from Table~\ref{tab:PTZs} (computed at
the same scale of $\mu=2\gev$, $\frac{Z_{DD,\partial\partial}}{Z_A}=0.027(8)$),
we arrive at the renormalised moments of the distribution amplitudes given in
Table~\ref{tab:final}.
\begin{table*}
  \caption{Final results in the chiral limit in \MSbar\/ at
    $\mu=2\gev$ for both of our lattice volumes. Here the first error
    is statistical, the second includes systematic errors from $m_s$,
    discretisation and renormalisation.}
  \label{tab:final}
  \begin{ruledtabular}
    \begin{tabular}{l|ccccccc}
                       & $\PSM_\pi$ & $\PFM_K$    & $\PSM_K$   & $\VSM_\rho$ & $\VFM_{K^*}$ & $\VSM_{K^*}$ & $\VSM_\phi$ \\
      \hline
      $16^3 \times 32$ & 0.25(1)(2) & 0.035(2)(2) & 0.25(1)(2) &  0.25(2)(2) & 0.037(1)(2)  & 0.25(1)(2)   & 0.24(1)(1)%
      \rule{0pt}{4mm}\\
      $24^3 \times 64$ & 0.28(1)(2) & 0.036(1)(2) & 0.26(1)(2) &  0.27(1)(2) & 0.043(2)(3)  & 0.25(2)(2)   & 0.25(2)(1)
    \end{tabular}
  \end{ruledtabular}
\end{table*}
The contribution from the mixing term in Eq.~\eqref{eq:resum2} is
small so using the perturbative result for $Z_{DD,\partial\partial}$
is not a drawback. Even if the correction of a nonperturbative result
for $Z_{DD,\partial\partial}$ is as sizeable as for
$Z_{\firstmomentop}$ or $Z_{DD,DD}$, the overall contribution remains
comparable to our present error on $Z_{DD,DD}/Z_A$. Hence our results
are essentially renormalised nonperturbatively.

\section{Summary}
\label{sec:conclusion}

We have computed the first or first two lowest non-vanishing moments
of the distribution amplitudes of the $\pi$, $K$, $K^*$, $\rho$ and
$\phi$ mesons, using nonperturbative renormalisation of the lattice
operators, with final numbers given in Table~\ref{tab:final}. Apart
from the uncertainty in $m_s$ for the first moments, systematic errors
mainly come from the renormalisation procedure. Within the current
statistical errors on our data we do not see any finite size effects.
With only one lattice spacing we can also only estimate a formal
discretisation error of $O(a^2 \Lambda_\text{QCD}^2)\approx 4\%$ from
the $O(a)$-improved DWF action and operators; this is included in our
sytematic error. The result for $\PFM_K$ in Table~\ref{tab:final}
supercedes but is compatible with our earlier result
in~\cite{Boyle:2006pw,Boyle:2006xq}, which was obtained on the
$16^3\times32$ ensembles only and used perturbative renormalisation.

Converting the lowest moment of the kaon distribution amplitude to the
first Gegenbauer moment $a_K^1=0.061(2)(4)$, we find it in agreement
with sum rule results from Eq.~\eqref{eq:other} but with a much
reduced uncertainty. We compare our results to those from the QCDSF
Collaboration~\cite{Braun:2006dg} in Table~\ref{tab:other}
(preliminary results for the first moment of the vector meson
distribution amplitudes are also available from
QCDSF~\cite{Braun:2007zr}). The results for $\PFM_K$ differ
significantly. However, we observe that our measurements correspond to
pion masses in the range $330$--$670\mev$ and are for $2+1$ dynamical
flavours, whereas the QCDSF results are for pion masses around
$600\mev$ and higher, with $2$ dynamical flavours. For one data point
from each collaboration where the pion and kaon masses are comparable,
the $\PFM_K$ values differ by about one standard deviation. These
points occur for the smallest values of $m_K^2-m_\pi^2$ from each
collaboration; for larger values the points, and therefore slopes in
$m_K^2-m_\pi^2$, differ.
\begin{table*}
  \caption{Comparison to other lattice results (both for \MSbar\/ at $\mu=2\gev$).}
  \label{tab:other}
  \begin{minipage}{.75\textwidth}
    \begin{ruledtabular}
      \begin{tabular}{l|cccc}
        & $\PSM_\pi$ & $\PFM_K$ & $\PSM_K$ & $\VFM_{K^*}$ \\
        \hline
        this work ($24^3\times 64$)& 0.28(1)(1) & 0.036(1)(2) & 0.26(1)(1) & 0.043(2)(3)%
        \rule{0pt}{4mm}\\
        QCDSF~\cite{Braun:2006dg} & 0.269(39) & 0.0272(5) & 0.260(6)
      \end{tabular}
    \end{ruledtabular}
  \end{minipage}
\end{table*}
We plan to improve our results in the near future by reducing the
systematic uncertainties. We will improve the nonperturbative
calculation of the renormalisation factors by including the total
derivative mixing term. We will also have an additional lattice
spacing allowing us to estimate the continuum results, including using
partially-twisted boundary conditions to remove hypercubic lattice
artefacts~\cite{Arthur:2010ht,Arthur:2010hy}. Increased statistics on
the $24^3\times64$ lattice should also improve our conclusions about
finite volume effects.

\begin{acknowledgments}
The calculations reported here used the QCDOC computers
\cite{Boyle:2005qc,Boyle:2003mj,Boyle:2005fb} at Edinburgh University,
Columbia University and Brookhaven National Laboratory (BNL). The
Edinburgh QCDOC system was funded by PPARC JIF grant
PPA/J/S/1998/00756 and operated through support from the Universities
of Edinburgh, Southampton and Wales Swansea, and from STFC grant
PP/E006965/1. At BNL, the QCDOC computers of the RIKEN-BNL Research
Center and the USQCD Collaboration were used. The software used
includes: Chroma~\cite{Edwards:2004sx}, QDP++ and the CPS QCD
codes~\cite{cps}, supported in part by the USDOE SciDAC program; the
BAGEL~\cite{Bagel} assembler kernel generator for many of the
high-performance optimized kernels; and the UKHadron codes.
We thank the University of Southampton for access to the Iridis
computer system used in the calculations of the nonperturbative
renormalisation factors (with support from STFC grant ST/H008888/1).
DB, MAD, JMF, AJ, TDR and CTCS acknowledge support from STFC Grant
ST/G000557/1 and from EU contract MRTN-CT-2006-035482 (Flavianet); RA
and PAB from STFC grants PP/D000238/1, PP/C503154/1 and ST/G000522/1;
PAB from an RCUK Fellowship.
\end{acknowledgments}

\bibliography{pda}

\end{document}